# HABITABLE ZONES IN THE UNIVERSE

GUILLERMO GONZALEZ

*Iowa State University, Department of Physics and Astronomy, Ames, Iowa 50011, USA*

Editorial correspondence to:

Dr. Guillermo Gonzalez

Iowa State University

12 physics

Ames, IA 50011-3160

Phone: 515-294-5630

Fax: 515-294-6027

E-mail address: gonzog@iastate.edu

**Abstract.** Habitability varies dramatically with location and time in the universe. This was recognized centuries ago, but it was only in the last few decades that astronomers began to systematize the study of habitability. The introduction of the concept of the habitable zone was key to progress in this area. The habitable zone concept was first applied to the space around a star, now called the Circumstellar Habitable Zone. Recently, other, vastly broader, habitable zones have been proposed. We review the historical development of the concept of habitable zones and the present state of the research. We also suggest ways to make progress on each of the habitable zones and to unify them into a single concept encompassing the entire universe.

# 1. Introduction

Since its introduction over four decades ago, the Circumstellar Habitable Zone (CHZ) concept has served to focus scientific discussions about habitability within planetary systems. Early studies simply defined the CHZ as that range of distances from the Sun that an Earth-like planet can maintain liquid water on its surface. Too close, and too much water enters the atmosphere, leading to a runaway greenhouse effect. Too far, and too much water freezes, leading to a runaway glaciation. Since these modest beginnings CHZ models have become more complex and realistic, mostly due to improvements in the treatment of energy transport in planetary atmospheres and the inclusion of the carbon-silicate cycle. Along the way, Mars and Venus have served as "real-world" test cases of the CHZ boundaries.

The CHZ has been an important unifying concept in astrobiology. To date, research on the CHZ has required knowledge of stellar evolution, planetary dynamics, climatology, biology, and geophysics. Yet, even modern CHZ models are far from complete. Many factors relating to planet formation processes and subsequent gravitational dynamics have yet to be incorporated in a formal way.

While they were not the first to discuss habitability beyond the Solar System, Gonzalez *et al.* (2001a,b) were the first to introduce a unifying concept called the "Galactic Habitable Zone" (GHZ). The GHZ describes habitability on the scale of the Milky Way Galaxy. While the GHZ appears superficially similar to the CHZ, it is based on a very different set of physical processes, including the radial gradients of the supernova rate and the gas metallicity in the disk. It should be possible to define habitable zones for other galaxies and to extend the concept to the whole universe. The largest of all habitable zones can be termed the "Cosmic Habitable Age" (CHA), which describes the evolution of the habitability of the universe.

In the following, published studies on the CHZ, GHZ, and CHA are briefly reviewed. Then, the deficiencies of the present definitions of habitable zones are noted and relevant physical processes that should be included in future studies are suggested. Gaps among the CHZ, GHZ, and CHA are also identified. Finally, ways to improve and unify the existing habitable zone concepts are suggested.

## 2. Published research on habitable zones

## 2.1 THE CIRCUMSTELLAR HABITABLE ZONE

All published studies of the circumstellar habitable zone (CHZ) start with an Earth-like planet. Here, "Earth-like" means a terrestrial planet with the same geophysics and initial inventory of volatiles as Earth. The planet is assumed to be habitable as long as liquid water can be maintained on its surface. It is imbedded in a planetary system identical to ours, except possibly a different host star. It has the same eccentricity, moon, and planetary neighbors. Thus, all the difficult questions about the formation of the Solar System are avoided. This is the traditional definition of the CHZ.

It may seem that the requirement of liquid water is merely an assumption of convenience for defining the CHZ based on our knowledge of "life as we know it." The evidence from chemistry, however, lends support to the view that liquid water is essential for life. Whewell (1833) already linked the habitability of the environment to the anomalous expansion of water on freezing. Henderson (1913) presented many more examples linking life to the anomalous properties of water, and he also presented similar arguments for carbon. The evidence for the essentiality of carbon and water for life has become stronger since Henderson's day (Barrow and Tipler ,1986; Lewis, 1998; Brack 2002).

Whewell (1853) was the first to propose that the Solar System has a habitable region comparable to the modern conception of the CHZ. He termed it the "Temperate Zone." In an impressive treatise for the period, Wallace (1903) enumerated several planetary habitability factors, including obliquity, mass, distance from the Sun, atmospheric composition, and proportion of water to land. Strughold (1955) termed the habitable zone around the Sun the "ecosphere." Huang (1959) described how the surface temperature of the Earth would vary if it were moved closer to or farther from the Sun. Dole's (1964) analysis was far more extensive and quantitative than any previous work, considering such factors as planet mass, rotation period, obliquity, and insolation variations. However, he did not explore the coupling among these factors, and some of his assumptions are dated. Later treatments of the CHZ retreated back to focus on climate evolution resulting from interactions among atmospheric, geochemical, and radiation transfer processes.

Hart (1978, 1979) presented a detailed and mathematical study of the CHZ. He modeled the evolution of the Earth's climate since its formation, including volcanic outgassing, atmospheric loss, the greenhouse effect, albedo variations, biomass variation, various geophysical processes, and the gradual brightening of the Sun. This was an early example of "Earth system" modeling and demonstrated that many processes must be treated simultaneously to properly account for feedbacks. Hart defined the inner and outer boundaries of his CHZ by runaway greenhouse and runaway glaciation, respectively. Based on these, he found a very narrow CHZ for the Solar System, ranging from 0.958 to 1.004 astronomical units (AUs).

Hart also introduced the continuously habitable zone concept, which takes the Sun's gradual brightening into account; as the Sun ages, its luminosity increases, causing the CHZ to migrate outwards. According to Hart, then, at any given time the continuously habitable zone is the region of overlap of all the previous CHZs. He also calculated the CHZ for other stellar spectral types. Hart's models marked the beginning of the modern era of CHZ studies.

Nevertheless, Hart was criticized for the way he treated cloud formation and the associated albedo variations. Cloud albedo variations are still not well understood, and models subsequent to Hart's have assumed constant surface and cloud albedo. In addition, Walker *et al.* (1981) showed that the temperature dependence of the weathering rate of exposed continental silicate rocks produces a stabilizing feedback on the surface temperature, which Hart had neglected in his models. Kasting *et al.* (1993) calculated a new set of CHZ models for the Sun and other stellar spectral types. They differed from Hart's models primarily by including the carbonate-silicate feedback cycle and a more accurate treatment of energy transport in the atmosphere.

Kasting *et al.* defined the inner boundary of the CHZ in multiple ways. One is based on the "moist greenhouse". In this process water gets into the stratosphere, where it is dissociated by solar UV radiation and the H atoms are lost from the top of the atmosphere. A second definition for the inner boundary is based on the runaway greenhouse. They calculated the outer boundary according to the maximum possible $CO_2$ greenhouse or the increase of planetary albedo due to formation of $CO_2$ clouds. The inner and outer boundaries were also estimated from the states of Venus and Mars, respectively. Their most restrictive case has inner and outer boundaries of 0.95 and 1.37 AUs, respectively.

Forget and Pierrehumbert (1997) and Mischna *et al.* (2000) studied the radiative effects of $CO_2$ clouds near the outer edge of the CHZ, and found that they may have a net warming effect. If the additional warming they cause by reflecting more infrared radiation is greater than the additional cooling they provide, then the outer boundary of the CHZ would be pushed outward. Of course, this only works for thick, predominantly $CO_2$ atmospheres.

Franck *et al.* (2001) presented a new set of CHZ models based on a more realistic treatment of geophysical processes. Previous studies had assumed constant continental area, metamorphic outgassing of $CO_2$, and weathering rate over geologic timescales. Building on the climate models of Kasting *et al.* (1993) and Caldeira and Kasting (1992) and relaxing these assumptions, Franck *et al.* thus modeled Earth's coupled climate-geologic systems as dynamical processes. Their CHZ is defined by surface temperature bounds of 0°C and 100°C and $CO_2$ partial pressure above $10^{-5}$ bar. They added the $CO_2$ partial pressure requirement to ensure that conditions are suitable for biological productivity via photosynthesis. It sets the inner boundary of their CHZ, while the minimum temperature requirement sets the outer boundary; their CHZ inner and outer bounds for the present Solar System are 0.95 and 1.2 AUs, respectively. Franck *et al.* also determined that the maximum lifespan for an Earth-like planet around a star between 0.6 and 1.1 $M_{sun}$ (6.5 Gyrs) is limited by planetary geodynamics.

Do these studies of the CHZ unreasonably restrict our focus to Earth-like planets? Pockets (or even oceans) of liquid water likely exist in the interiors of the giant planets, the icy Galilean Moons, Titan, and the Martian subsurface. Although these places are located outside the traditional CHZ, some argue that they should also be called habitable zones. Nevertheless, there are several reasons to keep the traditional definition. First, one of the primary motivations for study of the CHZ is to learn how to detect habitable terrestrial planets in other systems. Since we know Earth has been inhabited most of its history, finding another planet like Earth is probably the surest bet to find another inhabited planet. Second, since life requires much more than just liquid water, merely locating other pockets of liquid water will not suffice. Each new potential niche will have to be evaluated on its ability to supply the necessary ingredients for life in the right forms. It has yet to be demonstrated that these other niches can harbor life.

## 2.2 THE GALACTIC HABITABLE ZONE

Habitability on the scale of the Milky Way Galaxy has been discussed at least since Schklovsky and Sagan (1966) considered the perturbations to the biosphere by nearby supernovae (SNe).

Since then several studies have reexamined this question. Some have considered the effects of ionizing radiation from a nearby SN on Earth's ozone layer (e.g., Ruderman 1974; Gehrels *et al*. 2003). Others have searched the paleobiological and geological records for signatures of nearby SNe. For example, Benitez *et al*. (2002) present evidence linking the Pliocene-Pleistocene boundary marine extinction event about 2 Ma to a nearby SN, and Knie *et al*. (2004) and Wallner *et al*. (2004) discovered spikes in the $^{60}$Fe and $^{244}$Pu, respectively, concentrations in marine sediments from about the same time, consistent with a SN explosion about 40 parsecs from Earth.

SNe are not the only biologically important radiation events in the Galaxy. Clarke (1981) considered the possible biological consequences of our Galaxy undergoing a Seyfert-like nuclear outburst, and Scalo and Wheeler (2002) discussed the possible biological effects of a nearby gamma ray burst (GRB). There are also weaker sources of ionizing radiation (such as novae and cataclysmic variables), but their relevance to the biosphere has not yet been explored.

The Galactic environment also affects the orbits of comets in our Solar System and (presumably) others. The weakly bound Oort cloud comets are sensitive to large-scale gravitational perturbations, including the Galactic vertical (disk) (Heisler and Tremaine, 1986; Matese *et al*., 1995) and radial (Heisler and Tremaine, 1986; Matese *et al.,* 1999) tides, Giant Molecular Clouds (GMCs) (Hut and Tremaine, 1985), and nearby star encounters (Matese and Lissauer, 2002). Of these, the galactic disk tide is the dominant perturber of the present outer Oort cloud comets; the disk tide is about 15 times greater than the radial tide (Heisler and Tremaine). Matese *et al*. (1999) argue that the imprint of the radial Galactic tide is present in the observed distribution of long-period comet aphelia on the sky. Such perturbations can cause "comet showers" in the inner Solar System and thus increase the comet impact rate on Earth. However, empirical evidence tying the cratering record on Earth to Galactic perturbations remains controversial (Jetsu and Pelt, 2000).

Marochnik (1984) and Bal´azs (1988) argued that the placement of the Sun's Galactic orbit very near the corotation circle is an important requirement for habitability. Stars at the corotation circle in the disk orbit with the same period as the spiral arm pattern. Therefore, so the argument goes, the time interval between spiral arm crossings is greatest for a star with a circular orbit near the corotation circle, and such a star enjoys habitable conditions for longer duration than other stars not in such an orbit. Since massive stars are more common in the spiral arms, stars passing through them will be exposed to more nearby SNe. How close the Sun is to the corotation circle is still unresolved. Recent studies tend to fall into two camps; most, such as Lépine *et al.* (2001), Dias and Lépine (2005) and Fernandez *et al.* (2001), argue that the Sun is very near the corotation circle, while others (e.g., Martos *et al.*, 2004) argue it is not (see Shaviv 2003 for a summary of this debate).

Trimble (1997a,b) considered, in a general way, the metallicity of the Sun in the broader context of Galactic chemical evolution and how the metallicity of the gas in the interstellar medium constrains the formation of habitable planets. If the metallicity is too low, then it will not be possible to form an Earth-*like* planet, which is composed mostly of O, Mg, Si, and Fe. These elements are produced primarily by massive star SNe, which enrich the originally pure H and He interstellar gas with their processed ejecta. An Earth-*mass* planet can probably form from H, He, and water at lower metallicity, but it would probably be far less habitable (see below).

Tucker (1981) considered habitability in the Galaxy within a more general framework, but his treatment was superficial and is now seriously outdated. Gonzalez *et al.* (2001a,b) unified the various seemingly disparate Galactic factors by introducing the Galactic Habitable Zone (GHZ) concept. The inner boundary of the GHZ is set by the various hazards to the biosphere, and its outer boundary is set by the minimum metallicity required to form an Earth-like planet (Gonzalez, 2005). The perturbers include transient radiation events, comet impacts. The incidence of giant planets is much higher among metal-rich stars, but the planets tend to have highly eccentric or very short period orbits (Gonzalez, 2003; Santos *et al.*, 2004, 2005). Such orbits are less likely to be compatible with the presence habitable terrestrial planets (see below). The temporal evolution of the GHZ is determined primarily by the evolution of the metallicity of the interstellar gas, of the interstellar abundances of the geologically important radioisotopes ($^{40}$K, $^{235}$U, $^{238}$U, and $^{232}$Th), and of the rate of transient radiation events.

Lineweaver *et al.* (2004) further quantified the GHZ by applying Galactic chemical evolution models; they included the effects of the evolving interstellar gas metallicity and SN rate. They estimated that 10% of all the stars that have ever existed in the Milky Way have been in the GHZ. Peña-Cabrera and Durand-Manterola (2004) have also explored the GHZ, though less extensively.

2.3 COSMIC HABITABLE AGE

The Steady State Theory (Hoyle, 1948) implied that the universe has always appeared as it does today. The widespread acceptance of the Big Bang theory by the 1960s made it clear that such is not the case. Analysis of the Wilkinson Microwave Anisotropy Probe (WMAP) data indicates that the universe began in a hot dense state 13.7 Gyrs ago (Spergel *et al.*, 2003). Analyses of distant galaxies show that the global star formation rate has been declining for the past 5 Gyrs (Heavens *et al.*, 2004). Old stars in the Galaxy are observed to be systematically deficient in metals compared to young stars, mirroring the evolution of metals in the broader universe. The universe has changed drastically since its formation, and these changes bear on the question of habitability.

   Discussions of the "Cosmic Habitable Age" (CHA) have usually been framed in terms of the Anthropic Principle (Dicke, 1961; Garriga *et al.*, 2000; Rees, 2003). Given that the universe has changed so dramatically since its origin, the question naturally arises why we observe ourselves to be living during this particular time as opposed to some other time. Clearly, chemically based life is not possible in the very early universe before atoms formed or in the distant future, after all the stars burn out. Other considerations indicate that the boundaries of the CHA are narrower than these extreme limits.

   Lineweaver (2001) estimated the probability of forming Earth-like planets over the history of the universe based on the evolution of the global metallicity. He assumed that habitable terrestrial planets are most probable over a narrow range of metallicity centered on the solar value. von Bloh *et al.* (2003b) used the results from Lineweaver to calculate the peak time of the incidence of Earth-like planets in the Milky Way; they found it to be at about the time of formation of the Earth.

### 3. Refining the zones

Today, the CHZ, GHZ, and CHA are treated as separate concepts. Eventually, it should be possible to unify them into a single broad understanding of habitability for every galaxy and spanning the entire history of the universe. Before this goal can be achieved, it will be necessary to continue refining the habitable zone concepts within the existing frameworks. At the same time, it will be necessary to find the appropriate "hooks" to link these habitable zones together and fill in the gaps between them.

## 3.1 LEARNING FROM EARTH LIFE

The published studies of the CHZ described above focus on the maintenance of minimal habitable conditions on the surface of a terrestrial planet. These conditions are constrained most fundamentally by limits on the planet's mean surface temperature, the presence of liquid water, and the composition of its atmosphere. To these we can add constraints on the temporal and spatial variations of a planet's surface temperature; a slowly rotating Earth-like planet, for example, will experience larger temperature variations than a similar faster rotating planet having the same mean temperature.

Some astrobiologists have treated habitability as a binary, either-or quantity. A planet is either habitable or it isn't; it either has liquid water on its surface, or it doesn't. Frank *et al.* (2001) advanced beyond this simplistic approach; they quantified the habitability of a planet by calculating its photosynthetic productivity. While photosynthesis is not the most basic form of habitability, it is one that has existed on Earth since very early times. Following Franck *et al.*, we propose that a Basic Habitability Index (BHI) be adopted as a basic measure of habitability. We can additionally define a habitability index for Earthly animal life (i.e., large, oxygen-breathing, mobile metazoans); we can call it the Animal Habitability Index (AHI). According to the *Rare Earth* hypothesis (Ward and Brownlee 2000), the AHI would be more restrictive than the BHI. The limits on the mean surface temperature and the surface temperature variations would both be narrower for the AHI. An upper limit on the carbon dioxide partial pressure also needs to be added, as well as a lower limit on the oxygen partial pressure for the AHI. These limits can be estimated from the physiology of extant animals, the reconstructed evolution of the partial pressures of carbon dioxide and oxygen in Earth's atmosphere (e.g., Berner *et al.*, 2000), and the history of life. While such limits will be necessarily parochial, certain general physiological principles we have learned from Earthly life will apply universally.

The study of von Bloh *et al.* (2003a) is a first attempt to incorporate distinct types of biospheres in a long-term climate model. They modeled prokaryotes, eukaryotes, and complex multicellular life in a very simple way. Each type of life is defined by its own temperature tolerance window.

Particularly helpful in quantifying the AHI and BHI is knowledge of the global ecological patterns of the present Earth. Ecologists have noted that a few large-scale spatial patterns account for the distribution of biodiversity (Gaston, 2000). The most prominent among these are the decrease in biodiversity (quantified as species richness) with increasing latitude and altitude. More fundamentally, Allen *et al.* (2002) argue that biodiversity increases with increasing temperature and nutrient availability; they explain the temperature dependence in terms of basic biochemical kinetics. Biodiversity also correlates positively with primary ecosystem productivity (Waide *et al.,* 1999); for example, Schneider and Rey-Benayas (1994) show how the diversity of vascular plants correlates with productivity. Other factors that influence biodiversity and ecosystem productivity include temperature variability and mean insolation (for photosynthesis), both of which are more important at high latitudes. In order to create a complete model of the CHZ it will be necessary to include all these basic ecological factors.

Another possibly fruitful approach towards generalizing habitability would be to construct an "equation of state of life". For example, Méndez (2001, 2002) compiled a database of the physiological properties of several hundred genera of prokaryotes and studied statistical trends in it. Prokaryotes are an important element of the primary producers, and thus, of biodiversity. He found that about 85% of prokaryotes have an optimum growth temperature between 295 and 315 K. This is interesting, because it implies that the biophysical limitations of prokaryotes have been more important to their distribution on Earth than adaptations. A complete equation of state for prokaryotes would include at least temperature, pressure and water concentration as parameters.

The history of life on Earth is another important source of information on factors relevant to habitability (Nisbet and Sleep, 2001). The fossil record reveals that single-celled life appeared on Earth at least 3.5 Gyrs ago (Schopf *et al.*, 2002), shortly after the end of the "late heavy bombardment". The "Cambrian explosion" occurred about 540 Myrs ago. Since then, there have been many extinction events with global footprints (Sepkoski, 1995). Only the K/T extinction has been securely linked to a well-dated extraterrestrial event – the Chixzulub impact structure. Once additional extinction events can be linked securely to individual impacts, it will be possible to produce a "kill curve", which relates the magnitude of extinction and the size of the impact crater (Rampino, 1998). It will probably be necessary to include some threshold impactor energy required to trigger global extinctions given that other large impacts, such as the two that occurred 35.5 Myrs ago (Chesapeake and Popigai; about 100 km each), had relatively little global effect on the biosphere (Bottomley *et al.*, 1997). In addition, the damage inflicted by an impactor on the biosphere depends on additional factors, such as the composition of the impact site.

The volatile inventory of a terrestrial planet in the CHZ can vary dramatically depending on the details of its formation (see below). The habitability of a terrestrial planet depends sensitively on its total water content. Planets with much less surface water than Earth, like Mars, experience larger temporal and spatial temperature variations. On the other hand, planets with much more surface water are not necessarily more habitable. On first consideration we should expect such planets to have less variable surface temperature and therefore to be more habitable. However, reduced dry land area also means less opportunity for land-based life and less surface area for chemical weathering, an important part of the carbon-silicate cycle thermostat (Kump *et al.,* 1999). Marine organisms near the surface depend on nutrients and minerals washed off the continents and the regulation of the oceanic salt content by the continents. In addition, one type of origin of life scenario requires periodic drying (concentration) and wetting (dilution) of a solution containing biomolecules (Lathe, 2004); such a setting would not be available on a planet without dry land.

## 3.2 THE CIRCUMSTELLAR HABITABLE ZONE

### 3.2.1 SETTING THE INITIAL CONDITIONS

While CHZ models have improved steadily over the last few decades, they are still at an immature stage of development. They lack many deterministic and stochastic processes relevant to habitability, and the modelers have yet to describe how the formative processes of a planetary system set the initial conditions for their CHZ calculations.

The relevant initial conditions include the locations, masses, compositions, initial volatile inventories, initial rotation periods, initial obliquities, initial orbital inclinations, presence of moons, and initial eccentricities of the terrestrial planets and the orbits and masses of the giant planets; they also include the properties of the asteroid and comet reservoirs. These have significant stochastic components, and they cannot properly be treated in isolation, as there are many complex interdependencies among them. Proper treatment of the initial conditions requires simulations that begin with a protoplanetary nebula of a given mass, composition and environment and follow its evolution through the final stages of star and planet formation.

Planet formation is an active area of research, but real progress in theoretical modeling has only been possible over the past decade as computer processor speeds have increased dramatically and N-body codes have become more efficient. Lissauer (1993) identified four dynamical stages of planet formation in a gas and dust disk: 1) condensation and growth of grains, 2) grains grow to km size either by pairwise accretion or gravitational instability of the solid disk, 3) oligarchic growth to Mars-size terrestrial bodies and giant planet runaway accretion and 4) development of crossing orbits leading to giant impacts. Progress on the first two stages requires better knowledge of grain physics and treatment of non-gravitational forces. Non-gravitational forces can be largely ignored once the solid bodies reach about 1 km in size, but the vast numbers of such bodies (trillions) requires clever ways of doing the simulations efficiently (Barnes 2004). Numerical simulations have been most instructive on the late stages of planet formation. For example, Wetherill (1996) presented the results of gravitational simulations that followed several hundred planetesimals as they collided and eventually formed terrestrial planets (under the assumption that giant planets don't migrate). He found considerable diversity in the end states, which is not surprising since collisions are stochastic processes. Still, Wetherill and others (e.g., Raymond *et al.* 2004) have also shown that the final distributions of orbital periods, eccentricities, and masses of the terrestrial planets are significantly constrained. In all these calculations, though, the timescales of certain key processes are still uncertain. These include settling of grains to the mid-plane, giant planet formation, giant planet migration, and loss of disk gas.

The origin of planetary rotation is still unresolved. Simulations indicate that large impacts near the end of the accretion phase of forming terrestrial planets are likely to dominate any systematic preference for one spin direction over the other (Lissauer *et al.*, 2000). For example, the formation of the Moon via an impact by a Mars-size body probably imparted more angular momentum to the Earth than it had prior to that event (Canup, 2004). Following the early formative phase, the rotation periods of terrestrial planets continue to evolve via tidal torques from the host star and from any orbiting moons. Whether the rotation periods increase or decrease and how fast they change depend on the details of a planet's interior, oceans, and atmosphere as well as the direction of its rotation and the rotation period in comparison to its moon's orbital period.

Planetary rotation is highly relevant to habitability. A planet's rotation period affects its day-night temperature variation, obliquity stability (see below), and probably magnetic field generation. Unless a terrestrial planet has a thick carbon dioxide atmosphere, slower rotation will result in larger day-night temperature differences. In addition, prolonged absence of light will be a factor for photosynthetic life on any slowly rotating terrestrial planet. For the extreme case of synchronous rotation, the complete freeze-out of water on the dark hemisphere is very likely (see below). Once water begins to freeze on a region of a planet with continuously sub-zero temperature, the stage is set for a runaway process of continuing freeze-out.

The details of the origin of the atmospheres of the terrestrial planets are also uncertain. The two general classes of sources of volatiles are accretion from local material in the protoplanetary nebula and collisions with comets and small and large bodies from the asteroid belt. Among the volatiles, most research has focused on water, given its importance in defining the CHZ. According to protoplanetary disk models, Earth could not have received its water from material formed near 1 AU, as the protoplanetary disk temperature would have been too high for it to condense. Water must have been delivered from beyond about 2.5 AUs. Apparently, nearly all Earth's water came from large bodies in the region of the outer asteroid belt (Morbidelli *et al.*, 2000). Contrary to previous expectations, isotopic and dynamical data indicate that comets contributed no more than about 10% of Earth's crustal water (Morbidelli *et al.*).

The net quantity of water and other volatiles delivered to and retained by a terrestrial planet also depends on its size and location. Smaller planets, like Mars, are subject to a much greater degree of atmospheric impact erosion (Lunine *et al.*, 2003). Earth's gravity is sufficiently large so that impacts added much more to its atmosphere than they removed. Even the giant impact proposed to have formed the Moon probably removed only a modest portion of Earth's atmosphere (Genda and Abe, 2003, 2004; Melosh, 2003). The impact velocity depends, in part, on the impactor's original orbit and on the orbit of the impactee. Comets, which originate far from the Sun, will impact at higher velocity than objects from the asteroid belt. Likewise, terrestrial planets closer to their host star will encounter impactors at greater velocities. Higher velocity impacts tend to erode planetary atmospheres more effectively.

Lunine (2001) argued that the delivery of volatiles to the terrestrial planets in the Solar System should be very sensitive to the location and eccentricity of Jupiter's orbit. One of the critical quantities is the location of the innermost giant planet in relation to the so-called snowline. The presence of Jupiter near the snowline in the Solar System allowed it to transfer water-rich embryos efficiently from the asteroid belt into the terrestrial planet region. Recent N-body simulations of the formation of the terrestrial planets have generally confirmed this. Raymond *et al.* (2004) showed that increasing the eccentricity of Jupiter produces drier terrestrial planets, and moving it farther from the Sun produces more massive, water-rich planets; they also find that the volatile delivery has considerable stochastic variability.

The immediate effects of an impactor will depend on the water inventory on a planet's surface. In the case of Mars, the relatively dry surface will result in less of the impactor's energy carried around the planet by the impact-generated steam. This will also depend on the thickness of the atmosphere, but it is likely than larger planets will have denser atmospheres (see below). This implies that smaller planets tend to be drier and thus suffer less immediate global effects from impacts, but this needs to be confirmed.

Today, the radial distributions of asteroid and comet perihelia peak just outside the orbit of Mars (see Figure 1). As the outermost terrestrial planet, Mars takes the brunt of asteroid and comet impacts (except that its smaller size than Earth gives it a smaller cross section for collision).

**[Figure 1]**

### 3.1.3 LEARNING FROM OUR PLANETARY NEIGHBORS

With the discovery of the first extrasolar giant planet around a nearby Sun-like star in 1995, it became immediately obvious that other planetary systems can be very different from ours. About 10% of the detected systems have a giant planet within about 0.1 AU of their host stars. The remaining systems have giant planets with eccentricities that scatter nearly uniformly between 0.0 and 0.80. New processes were proposed to account for the great variety of orbits observed. These include inward planet migration (Lin *et al.*, 1996; originally proposed in the 1980s) and strong disk-planet and planet-planet interactions (Chiang *et al.*, 2002; Goldreich and Sari, 2003; Marzari and Weidenschilling, 2002). Some of these processes also result in non-coplanar orbits, which tend to produce less stable systems (Thommes and Lissauer, 2003). Veras and Armitage (2005), assuming that the observed eccentricities are due to planet-planet scattering, determined that terrestrial planets are unlikely to form in a star's habitable zone if an eccentric giant planet has a semimajor axis between 2 and 3 AUs.

Ward (1997) identified two types of inward planet migration resulting from gravitational torques between a growing protoplanet and the surrounding disk. Migration is linear with mass in type I migration. If a planet grows to sufficient mass, it opens a gap in the disk, resulting in type II migration; in this case the migration rate is determined by the viscosity in the disk and not by the planet's mass. Current models of type I migration are much too efficient; the amount of type I migration must be reduced by a factor of 10 if planets are to be presented from spiraling into their host stars (Alibert *et al.* 2005). Thommes and Lissauer (2005) also discuss Type III migration, wherein the giant planet's mass is comparable to the mass of the disk. In this case, the migration rate is inversely proportional to the planet's mass, and it slows the evolution of the disk. Since planet formation, disk evolution, and migration occur on similar timescales, all three must be included in simulations for self-consistency.

Several studies have sought to determine if terrestrial planets can exist in the CHZs of known extrasolar planetary systems, given the gravitational perturbations from the giant planets in them (Menou and Tabachnik, 2003; Jones and Sleep, 2002). Menou and Tabachnik also recognize the limitations of current definitions of the CHZ, which assume stable circular orbits, and introduce the concept of "dynamical habitability." By this they mean whether a terrestrial planet can remain in the CHZ in the presence of gravitational perturbations from the giant planets in the system.

Chambers and Wetherill (2001) showed that the survival of the asteroid belt and the delivery of volatiles from it to the terrestrial planets are very sensitive to the properties of the giant planets. Wetherill (1994) showed that the comet flux into the terrestrial planet region of the Solar System is also very sensitive to the properties of the giant planets.

Giant planet migration is also important to the habitability of the terrestrial planets. For example, migration of a giant planet toward its host star will remove any terrestrial protoplanets in the CHZ and can reduce the probability that more planets will form there afterward (Armitage 2003). However, if a planet migrates quickly (within the first Myr of the disk's lifetime), then the solid bodies in the CHZ region will not be strongly depleted. On the other hand, migration is probably not limited to one giant planet per system. If the conditions are such that one giant planet forms and migrates quickly, then other giant planets in the system are likely to migrate as well, more effectively disrupting any terrestrial planets. This is an area requiring much more research, as the effects of migrating giant planets are complex and sensitive to the details.

While giant planet migration sometimes prevents habitable terrestrial planet formation, some have argued that it opens up new niches for other kinds of habitats. Kuchner (2003) and Leger *et al.* (2004), for example, note that an icy planet like Uranus or Neptune or something smaller, if it migrates into the CHZ, can become an "ocean planet." Such a planet would have a very deep ocean of water surrounding a thick ice mantle, which would separate the deeper silicate mantle from the ocean. The pressure at the bottom of its ocean would be too high for any known life. Such a planet would also be more sensitive to tidal torques from it host star and any large moons, causing more rapid spin-down. A small influx of life-essential elements at the surface could be provided by micrometeorites, but the quantities could probably only support at most a feeble biosphere.

Migrating giant planets would probably bring along at least some of their moons. How habitable would an Earth-size moon in the CHZ be? Williams *et al.* (1997) explored this possibility. Even if such a moon could be as large as Earth, it is unlikely to be as habitable for several reasons (Gonzalez *et al.* 2001a). Like its giant planet host, a moon formed far beyond the CHZ would contain a great deal of water; a possible exception would be a large moon intermediate in composition between Io and Europa. Other relevant factors include rotational synchronization (causing longer days and nights), tidal-induced migration, immersion in the host planet's radiation belts, and the higher frequency and energy of small-body impacts (e.g., the comet Shoemaker-Levy 9 impact and crater chains on Ganymede and Callisto; Schenk *et al.* 1996).

Lorentz *et al.* (1997) considered another kind of habitable moon. They predict that Titan might achieve sufficiently warm surface temperature during the Sun's red giant stage to be habitable for a few hundred million years, until the Sun strips away its atmosphere. Of course, the quality of the light from the future Sun will be very different from the present Sun, producing relatively less blue light (see discussion below about photosynthesis).

Dvorak *et al.* (2004) examine the stability of terrestrial planets in Lagrangian equilibrium $L_4$ and $L_5$ points, analogous to Jupiter's Trojan asteroids. They find that the sizes of these stable regions are sensitive to the eccentricity of the giant planet, being smaller for more eccentric systems. How such a "Trojan planet" could form at such a location or remain there during giant planet migration are issues needing additional research. Additional considerations include the composition of planets formed in proximity to a giant planet.

The giant planets also have significant influences on the obliquity variations of the terrestrial planets. Laskar *et al*. (1993) showed that a terrestrial planet can exhibit large and chaotic obliquity variations, which are caused by resonances between its precession frequency and combinations of secular orbital frequencies of the planets in the system. The chaotic zones are broad, and they depend on several parameters, including the orbital period, rotation period and mass of the terrestrial planet, the presence of a large moon and the orbital periods and masses of the giant planets. Certain combinations of these parameters result in very small obliquity variations. Today, Earth exhibits obliquity variations of only ±1.3 degrees around an average value of 23.4 degrees. The lunar gravitational torque increases Earth's precession frequency by a factor of about three compared to what it would be without the Moon, taking it far from a spin-orbit resonance (Laskar and Robutel, 1993). The Moon has a similar effect to reducing Earth's rotation period by the same factor.

Ward *et al*. (2002) demonstrated that the region of chaotic obliquity variation is very broad in the Solar System. They calculated the amplitude of obliquity variations for Mars over a broad range of locations and rotation periods. They found stability comparable to Earth's only for distances less than 0.7 AU from the Sun and with faster rotation (but solar-induced tides would slow the rotation over a few billion years). Interestingly, if Mars had a large moon (keeping all else the same), it would still exhibit large obliquity variations, because it would be brought closer to a resonance (see Figure 1 of Ward *et al*.).

A moonless Earth would have exhibited a stable obliquity if its rotation period were less than about 10 hours. The likelihood of such a state depends primarily on the last few large collisions it experienced near the end of its formation. Earth's initial rotation period was indeed less than 10 hours, but it has since slowed to 24 hours mostly by the action of the lunar tides. Ironically, Earth likely received its fast initial rotation from the impact that resulted in the Moon's formation. Thus, Earth's rotation is intimately tied to the presence of the Moon. Earth's obliquity variations would also be lessened if Jupiter were located at 3 AU instead of its present 5.2 AU. However, as noted above, such a configuration probably would have resulted in the delivery of fewer volatiles to Earth, and it might have prevented its formation in a stable orbit.

Mercury is presently locked into a 3:2 spin-orbit resonance with a stable low obliquity, but it was very likely born with a much faster rotation. Its precession frequency gradually declined via core-mantle interactions and tidal dissipation from the Sun that gradually slowed its rotation. Before reaching its present state, Mercury passed through a large chaotic zone in obliquity.

The case of Venus offers additional insights on obliquity variations. Correia *et al*. (2003) and Correia and Laskar (2003) show that most initial conditions drive Venus to its present state and that this is generally true of terrestrial planets with thick atmospheres. What's more, as long as Venus has a thick atmosphere, it avoids a 1:1 synchronous rotation. According to their simulations, then, Venus did not have to begin with a retrograde rotation to end up with one today. This tends to lend support to those who argue that terrestrial planets are formed predominantly with prograde rotation.

Touma and Wisdom (2001) studied the core-mantle, spin-orbit interactions for Earth and Venus and concluded that both planets have passed through major heat-generating core-mantle resonances. They speculate that Earth's passage through such a resonance about 250 Myr ago may have been responsible for generating the Siberian traps and causing the Permo-Triassic extinction and that Venus' passage through a similar resonance caused the planet to resurface itself about 700 Myr ago and generate its thick atmosphere. They find that terrestrial planets with retrograde rotation generate much more heat from such resonance passages. In addition, the Moon's tidal torque on Earth allowed it to pass quickly through its resonance, avoiding the fate of Venus. They also speculate that Venus' high surface temperature caused by its thick atmosphere maintained a magma ocean, which led to a rapid slowdown of its rotation through tidal dissipation. As a result, Touma and Wisdom argue that Venus was born with retrograde rotation; otherwise, it would have generated less heat through its core-mantle resonance passage. The case of Venus shows us how intimately linked are the geology, rotation, obliquity, orbit, and atmosphere of a terrestrial planet.

Both the value of the obliquity and the amplitude of its variation affect the habitability of a planet. Seasonal variations would be absent on a planet with a stable obliquity near zero degrees. While it would have constant surface temperatures, this benefit to life would be offset by two problems. First, weather systems would be more constant, some areas receiving steady precipitation, others receiving very little. More seriously, the Polar Regions will experience smaller maximum surface temperatures. Analogous to the water "cold traps" on synchronously rotating planets, polar ice would extend to lower latitudes, and it is possible that all the water would eventually freeze-out at high latitudes. A thick atmosphere would be a possible way out of such a state, as would a deep ocean, but they would have other consequences for life that would have to be examined in detail.

At the other extreme, a stable obliquity near 90 degrees would result in very large surface temperature variations over most of the surface of a terrestrial planet. Most planets will have unstable obliquities over at least a few Gyr, varying between small and large angles over millions of years. Each case will have to be treated in detail to determine overall habitability. Williams and Pollard (2003) have explored seasonal surface temperature variations for a wide range of obliquities using general circulation climate models of Earth-like planets, confirming that high obliquities produce more extreme variations in surface temperatures.

It is interesting that Earth's Moon has likely played several important roles in the history of life (as noted above): strong ocean tides possibly aiding in early prebiotic molecule formation, slowing Earth's rotation, stabilizing Earth's obliquity, reducing the heat generated during a core-mantle spin-orbit resonance and tidal mixing of minerals from the continents to the oceans. Recently, it has also been demonstrated that a significant fraction of the energy that drives the ocean currents comes from dissipation of lunar tidal energy in the deep oceans (Egbert and Ray, 2000; Wunsch, 2000). This is important, as the oceans are important to transferring heat from the equator to the Polar Regions. As the probability of the formation of a moon with the properties needed for its life-assistance is likely small, this could be a major limiting factor on the number of habitable planets in the universe (Waltham 2004).

Low eccentricities characterize the orbits of the planets in the Solar System. The present eccentricity of Earth's orbit is 0.016, smaller than most of the other planets (Earth's long-term mean eccentricity is even smaller than that of Venus, which presently has a smaller eccentricity than Earth). While exhibiting chaos, the planets have maintained low eccentricity orbits since they formed (Laskar, 1994). The Solar System is still the only planetary system for which we have a complete inventory of planets. For a system wherein only a fraction of the major planets are known, all the significant perturbations on any possible terrestrial planets in the CHZ cannot be determined. For this reason, long-term numerical simulations of the dynamics of planets are still tentative for extrasolar planetary systems. Numerical experiments with the Solar System can tell us how sensitive it is to individual changes.

We have conducted a number of numerical gravitational simulations using a second-order mixed-variable symplectic algorithm in MERCURY (v. 6.2; Chambers, 1999). In each run Earth, Venus, and the four giant planets were integrated for 15 Myrs, starting at the present (the small planets Mercury, Mars, and Pluto were excluded in order to reduce the computation time). The duration of the integrations was selected so each case would run in less than a day with the computer available for this project. Two of the cases were also integrated for 45 Myrs.

The most important results are listed in Table 1. Earth's eccentricity increases significantly as Jupiter's orbit is made more eccentric; its orbit becomes unstable if Jupiter's eccentricity is greater than about 0.15. Increasing the mass of Jupiter reduces the eccentricity of Earth's orbit, while reducing its mass increases it. Decreasing Jupiter's semi-major axis has a significant effect on Earth's orbit only near resonances. Earth's orbit becomes significantly more eccentric if its mass is less than about half its present value. Interestingly, in every case the mean eccentricity of Venus' orbit is greater than Earth's. The two longer integrations gave very similar results to shorter ones. Still, these cases underestimate the maximum eccentricity reached given their short duration. This will be especially important near resonances.

**[Table 1]**

Once it is known how the eccentricity of a terrestrial planet evolves, it is possible to calculate how its climate responds. There are two ways to do this for Earth. Williams and Pollard (2002) approached the problem with simulations of the climates of Earth-like planets under the assumptions of different eccentricities. Not surprisingly, they found that larger eccentricities produce larger annual temperature variations; larger eccentricity can also reduce the amount of time a planet spends in the CHZ. One of the most important factors describing the response of surface temperature to insolation variation is the radiative time constant of the atmosphere. It depends on, among other quantities, the surface pressure and heat capacity of the atmosphere. Earth's atmosphere has a time constant of about one month; planets with thicker atmospheres will have longer time constants. If the radiative time constant is much smaller than the orbital period, then a planet's surface temperature will be more sensitive to eccentricity-induced insolation variations.

Study of Earth's ancient climate via proxies stored in sediments and polar ice is another way to elucidate the relationships between orbital variations and climate change. The Milankovitch cycles (obliquity, precession, and eccentricity) have been detected in several such records (Berger and Loutre, 1994; Petit *et al.,* 1999; EPICA, 2004). Even the small ranges of variation in the obliquity and eccentricity of $23.4 \pm 1.3$ degrees and $0.00 - 0.04$, respectively, have been sufficient to leave their marks in the paleoclimate records. Earth's climate has been fluctuating dramatically in response to these small forcings for the past three million years. The connections between variations in Earth's orbital parameters and its climate are complex and much remains to be learned (Paillard, 2001).

## 3.1.4 PLANETARY GEOPHYSICS

The size of a terrestrial planet affects its habitability in diverse ways, yet few studies have explored these relationships explicitly. Hart (1982) and Kasting *et al.* (1993) considered the effects of changing the size of a terrestrial planet on the evolution of its atmosphere. Although the same criticisms noted above about Hart's other work apply here, his study is still helpful. Hart described how outgassing, atmospheric escape, and surface chemical processes depend on the size of a terrestrial planet. Kasting *et al.* noted that the greenhouse effect, albedo, atmospheric loss, and internal heat flow have substantial dependencies on a planet's size. Also, Lewis (1998) and Lissauer (1999) noted that a larger terrestrial planet, all else being equal, should have a deeper ocean and higher surface pressure due, in part, to the increasing importance of self-compression for terrestrial planets larger than Earth.

A planet's geophysics is also sensitive to its size. A smaller planet will lose its interior heat more quickly, primarily due to its larger surface area to volume ratio. A smaller planet also has smaller pressure throughout its interior, which affects core formation (Agee, 2004). Mars is an important comparison case; it is half the size of Earth and lacks plate tectonics and a global magnetic field. Evidence indicates that it did once possess a global magnetic field and was volcanically active for about the first billion years (Zuber 2001). The generation of a global magnetic field is closely linked to the operation of plate tectonics (Nimmo and Stevenson, 2000). Plate tectonics produces a larger surface heat flux than a one-plate mode of tectonics, which, in turn, produces a larger temperature gradient in the deep interior and convection in the liquid portion of the core.

A planetary magnetic dynamo requires a thermally or compositionally driven convecting fluid region of the metallic core (Stevenson, 2003). The presence of alloying light elements affects the melting temperature of iron and can cause chemical convection in the core (Stevenson). In the case of Mars, the volatile element sulfur is speculated to be the most abundant light element in its core (Spohn *et al.,* 2001). In the early stages of its growth a terrestrial planet's feeding zone spans a relatively narrow range in the protoplanetary nebula (Lissauer, 1995). It is only during the later stages of its growth that a terrestrial planet accretes embryos from more distant regions. Thus, because of the negative radial temperature gradient in the protoplanetary disk, terrestrial planets farther from the Sun should have more sulfur in their cores. All these theories are still preliminary and require considerably more research.

Very little research has been done on the biological consequences of a weak or absent magnetic field. It has been speculated that Earth's field is weakened sufficiently during polarity reversals to cause extinctions, though no such evidence has yet been found (e.g., Biernat *et al.*, 2002). Better established is the effect of the absence of a magnetic on a planet's atmosphere. The absence of a strong magnetic field over most of Mars' history has been implicated in the loss of a substantial fraction of its atmosphere through solar wind stripping (Jakosky and Phillips, 2001; Chassefière and Leblanc, 2004). The absence of a magnetic field for Venus has also been implicated in the loss of its water (Chassefière, 1997). An additional consequence might be increased secondary cosmic ray particle radiation at the surface of a planet, but this needs to be confirmed.

The exchange of water between a planet's interior and its surface also depends on its geophysics (Hauck and Phillips, 2002; Franck *et al.,* 2001). Over billions of years water on the surface of a terrestrial planet is lost to space and sequestered into its mantle (Bounama *et al.*, 2002). Water itself is intimately linked to the operation of plate tectonics (Regenauer-Lieb *et al.*, 2001). It is likely that the present mode of Venus' geophysics depends, in part, on the loss of its water.

3.1.5 THE HOST STAR

A planet's host star plays very important roles in the evolution of the CHZ. The host star affects the planets with its gravity and radiation. Wetherill (1996) ran planet formation simulations for a range of stellar masses. He found that the place of formation of terrestrial planets is relatively insensitive to the mass of the host star. This implies that terrestrial planets will tend to form outside the CHZ for stars less massive than the Sun and inside the CHZ for those that are more massive. Assuming the core-accretion model for giant planet formation (Pollack *et al.* 1996), Laughlin *et al.* (2004) argued that giant planets are less likely to form around stars less massive than the Sun; this theoretical result is consistent with the preliminary observation of Laws *et al.* (2003) that giant planets are less common around low mass stars. Without giant planets, the formation of terrestrial planets and the distributions of small bodies will be very different from those in the Solar System.

The Sun's radiation keeps water liquid on Earth's surface and allows plants to produce chemical energy from photosynthesis, but it also has negative effects on life. The insolation at Earth varies over many timescales. There is evidence that variation in the Sun's activity from decades to millennia affects Earth's climate (see Section 3.2.1). From stellar evolution models we know that the Sun has brightened by about 30% since its formation 4.6 Gyrs ago; this is taken into account in modern CHZ models.

We do not yet understand the Sun's chromospheric activity well enough to calculate how it has evolved since its formation. We can infer the evolution of the Sun's activity from observations of nearby Sun-like stars spanning a broad range in age. Such a research program, called "The Sun in Time", was begun about 20 years ago (Ribas *et al.*, 2005). Young Sun-like stars are observed to have shorter rotation periods, higher UV and X-ray luminosities, and more frequent flares (which produce temporary high fluxes of ionizing radiation). From satellite observations of such stars, the Sun's X-ray luminosity is inferred to have decreased by about 3 orders of magnitude, while its UV declined by about a factor of 20 (Güdel, 2003; Ribas *et al.*). Optical observations also imply that the Sun's optical variability has declined by a factor of about 50 over the same period (Radick *et al.*, 2004). Interestingly, Radick *et al.* also confirmed that the Sun's optical variability on decadal timescales is anomalously small compared to otherwise similar stars. The early higher activity is highly relevant to habitability, in part, because the higher flux of ionizing radiation stripped a significant fraction of the terrestrial planets' atmospheres (Pepin, 1997).

Chromospheric activity also correlates with the rate of flares. Flares produce X-ray and proton radiation. X-rays entering the top of the atmosphere are downgraded to UV line emission at the bottom of the atmosphere (Smith *et al.* 2004). Protons can alter the chemistry in the middle atmosphere and stratosphere. In particular, the two strongest solar proton events of the past four decades (1972 and 1989) were calculated to have produced reductions of the total ozone of 1-2 percent (Jackman *et al.* 2000). Even stronger flares should occur over longer timescales, and they should have been more frequent in the Sun's past (see Smith *et al.* for the flare frequency-energy relation).

Stellar activity also varies along the main sequence. Many red dwarf stars exhibit extremely strong and frequent flares. West *et al.* (2004) studied a large sample of nearby M dwarfs and found that activity peaks near spectral type M8 at about 80% incidence. The UV flux can increase by a factor of 100 during a flare (Gershberg *et al.*, 1999). Active M dwarfs have a soft X-ray to bolometric flux ratio several orders of magnitude greater than the same ratio for the Sun; during flares this flux ratio can be $10^6$ times the Sun's. Since the size of the CHZ is set by the bolometric flux, a planet in the CHZ of an M dwarf will be subjected to a much greater flux of ionizing radiation. Smith *et al.* found that the UV flux in the lower atmosphere of a planet around an M dwarf is completely dominated by redistributed energy from flare X-rays and reaches biologically significant levels.

The red color of an M dwarf star means that relatively less blue light will reach the surface of its orbiting planets compared to the Sun. Although photosynthesis doesn't require blue light, it generally becomes less effective without abundant light blueward of 6800 Å. Wolstencroft and Raven (2002) showed that Earth-like planets in the CHZ of cooler stars should be less effective at producing oxygen from photosynthesis. Some bacteria can still use infrared light, but not to produce oxygen. Any marine photosynthetic organisms would have difficulty using red light as an energy source, since ocean water transmits blue-green light much better than violet-blue or red light. The precise wavelength of peak transmission will depend on the minerals dissolved in the oceans. The transmittance of pure water peaks at 4300 Å, while that of Earth's oceans peaks a few hundred Å to the red.

A reduced stellar UV flux would slow the rate of photodissociation of abundant hydrogen-containing molecules such as water and methane in a planetary atmosphere. One of the primary ways to build up oxygen is to lose hydrogen atoms resulting from photolysis of such molecules (Claire *et al.*, 2004). It is likely that this process will be slower on a terrestrial planet orbiting an M dwarf, but this needs to be confirmed that the overall dissociating UV flux in such a case is smaller than for the Earth.

Planets around M dwarfs are often considered to be less habitable than Earth also because they achieve rotational synchronization too quickly (Kasting *et al.*, 1993). If a rotationally synchronized planet has a circular orbit, then one side will continuously face its host star while the other remains in darkness. This will lead to large temperature differences between the day and night sides and to freeze out of its water and possibly of other volatiles. Rotational synchronization can be avoided in two ways. First, a planet can have a sufficiently eccentric orbit so that, like Mercury, it has a 3:2 spin-orbit resonance. Such a planet would avoid water freeze-out, but it would still exhibit large temperature variations over the course of its orbit. The second way is to have a planet-size moon orbiting around a giant planet in the habitable zone. Such a planet, if it could exist in such a configuration, would suffer from many of the same difficulties already noted above.

Joshi *et al.* (1997) and Heath *et al.* (1999) produced simple climate model simulations of rotationally synchronized Earth-like planet around an M dwarf; the cases included surface pressures of 0.1, 1, and 1.5 bar of $CO_2$. They concluded that the higher surface pressure cases can exchange sufficient heat between the lit and dark sides to keep liquid water over much of its surface. Joshi (2003) ran three new cases with a more advanced atmospheric circulation model (all with 1 bar of $CO_2$): 1) a waterworld, 2) a northern hemisphere continent, and 3) dry land everywhere. In all three cases, temperatures low enough for ice formation occur on the dark side. Thus, water ice forms on the dark side, even with a pure $CO_2$ atmosphere with the same surface pressure as Earth's atmosphere.

This is not a problem for the waterworld case, since sea-ice will not change the sea-level and would remain on the dark side. In any case, as we noted above, a waterworld with deep oceans probably has a very low habitability. If there is some dry land or shallow seas on the dark side, the inevitable result is complete freeze-out of the water on the dark side. Once water-ice begins forming on land, the ocean level will begin to drop around the planet, exposing more land where stable ice can form, lowering the water level more, exposing more land, and so on until all the water freezes. In addition, as the water level drops, the oceans will become less efficient at transporting heat between the hot and cold regions, further accelerating the water freeze out. The planet will grow hot and dry on the lit side and cold and dry on the dark side. Some liquid water may remain at the bottom of some of the basins on the dark side, warmed by heat escaping from the planet's interior, but that will diminish as a planet ages. Even some elevated land near the terminator (such as mountain ranges) can trigger a runaway planetary freeze-up as long as ice can form at altitude. Even shallow seas on the dark side would allow for ice to accumulate and the sea level to drop as long as the ice can rest on the sea bottom. In short, surface relief will tend to destabilize a rotationally synchronized planet towards the complete freezing out of its water on its dark side. Another major destabilization point is reached if the dropping temperature on the dark side allows carbon dioxide to begin freezing out; this would eventually cause freeze-out of the entire atmosphere. Again, this is more likely to occur if there is significant surface relief. One needs a suspiciously contrived planet, with only continents on the lit side and deep oceans everywhere else, to prevent freeze-out of its water. But, as soon as continued tectonics and volcanic activity form land on its dark side, the entire planet will be set on its way towards the freeze-out of all its water. Even with an initially thicker carbon dioxide atmosphere, removal of carbon dioxide via the deposition of carbonates on the ocean floor would also cool the planet.

In summary, the definition of the CHZ depends on much more than just a planet's distance from its host star. A planet's habitability also depends on its orbital eccentricity, presence of a large moon, size, initial volatile inventory, initial rotation period and its evolution, the locations and properties of any giant planets, and the number and locations of small bodies. Much progress is being made on each of these factors, but eventually they will have to be treated together, given their sometimes strong interdependencies. Given the initial conditions established by the larger Galactic environment, CHZ models will have to start with a protoplanetary disk consisting of gas and dust and follow its evolution for billions of years. Due to the stochastic nature of many of the processes, thousands of simulations will be required to obtain useful statistics on the net habitability probabilities.

## 3.2 THE GALACTIC HABITABLE ZONE

The larger Galactic environment includes the immediate stellar and interstellar environment and the metallicity of the birth cloud, which depends on the location and time of its formation. Most disk stars form in star clusters. Star clusters form in various environments. Some, like the Taurus association, contain relatively few massive stars, while others, like Orion, contain numerous massive stars. Massive stars alter their environment via their ionizing radiation and by exploding as supernovae. They can partially photoevaporate protoplanetary disks (Störzer and Hollenbach, 1999) and seed them with short-lived radioisotopes (Lin *et al*. 2005). Close encounters by the more numerous intermediate and low mass stars in a cluster can also disrupt or make more eccentric the orbits of the outer planets (Laughlin and Adams, 1998; Adams and Laughlin, 2001). Unless nearby star encounters and the ionizing radiation and short-lived radioactive nuclei produced by massive stars are somehow required to form a habitable planetary system, low stellar density seems to be the preferred formation environment.

Long after a star leaves its birth cluster, Galactic-scale processes continue to influence life on a planet. These include transient radiation events, comet impacts, the background Galactic cosmic ray (GCR) flux, and passages through GMCs. As noted above, there has been progress towards understanding the possible biological effects of transient radiation events. Note, in the following discussion it is explicitly assumed that transient radiation events and other large environmental fluctuations are harmful to a biosphere, though some have suggested that they can be beneficial in some ways (e.g., Smith *et al*. 2004).

## 3.2.1 THE GCR-CLIMATE CONNECTION

Eddy (1977) presented the first convincing evidence for extraterrestrial influence of Earth's climate. He showed that the $^{14}C$ production rate in the atmosphere correlates with the sunspot cycle and with the climate of the past few centuries. Other studies have since found additional correlations on longer timescales. Bond *et al*. (2001) employed $^{14}C$ and $^{10}Be$ as proxies for solar variations and marine sediments in the North Atlantic as proxies for polar ice extent. They found a strong correlation between the polar climate and variations of solar activity on centennial to millennial timescales. Hu *et al*. (2003) also found correlations between the Sun and climate in lake sediments from southwestern Alaska. These two studies imply that the Northern Hemisphere climate varied in response to solar variations over the entire Holocene.

Several mechanisms for solar-induced climate change have been proposed. One involves the modulation of the GCR flux by the solar wind, which varies in strength over the sunspot cycle. Marsh and Svensmark (2000) presented evidence for a correlation between the global low cloud cover and the GCR flux between 1983 and 1994. However, there is some controversy as to whether the lack of a correlation since 1994 is an instrumental artifact or a real breakdown in the correlation (Marsh and Svensmark 2003a,b; Haigh 2003).

The implications of a possible GCR-cloud link are far reaching. Not only does the Sun modulate the flux of GCR reaching Earth, but so do fluctuations in Earth's magnetic field strength. In addition, the GCR flux from Galactic sources varies over long timescales. Thus, the importance of GCR-induced low cloud formation will depend on the Galactic star formation rate and on the location of the Solar System in the Galaxy (see next section). It also implies that the carbon-silicate cycle is not as important for long-term climate regulation as it is currently believed to be. For example, Wallmann (2004) modeled the Phaneozoic climate and found that the GCR-cloud link is needed to explain the observed oxygen isotope record in addition to the carbon-silicate feedback.

## 3.2.2 TRANSIENT RADIATION EVENTS

Transient radiation events important on the Galactic scale include supernovae (SNe), Gamma Ray Bursts (GRBs), Soft Gamma-ray Repeaters (SGRs), and Active Galactic Nucleus (AGN) outbursts. Less powerful events, such as novae, while more frequent, are less important on average and will not be considered below. Transient radiation events from stellar flares, while important, will not be discussed in this section (see Section 3.1.5). The rates, distributions, and energies of these four classes of Galactic transient radiation events will be briefly reviewed along with discussions of their possible effects on the biosphere.

The possible threats posed by SNe to life on a particular planet over the history of the Milky Way Galaxy depend on the spatial distribution, temporal frequency, and total radiant energy of the SNe. Galactic chemical evolution models are required to estimate both of the spatial distribution and temporal evolution of SNe, but observations can gives us part of the answer. A simplifying starting assumption is that the average SN rate in galaxies similar to the Milky Way (Hubble type Sbc) is representative of the rate in the Milky Way (Dragicevich *et al.*, 1999). This is probably a good assumption, but over relatively short time intervals the Milky Way's SN rate will sometimes deviate from the average significantly. Extragalactic SN surveys also yield reliable estimates of the average rates (Capellaro, 2004) of the various types of SNe (the main types are Type Ia (SN Ia) and Type II+Ib/c (SN II)) and their absolute magnitude distributions (Richardson *et al.,* 2002). Supernova rates are usually given in units of number per century per $10^{10}$ solar luminosities of blue light (SNu). For example, the total SN rate for elliptical galaxies is $0.18 \pm 0.06$ SNu, increasing to $1.21 \pm 0.37$ SNu for Scd-Sd Hubble types (Capellaro); SNIa are the only type of supernovae observed in elliptical galaxies. The total SN rate in the Milky Way is 2-3 SN per century (Ferrière 2001).

It is helpful to know the rates for the SN Ia and SN II separately, given their different distributions in the Galaxy. SN II, which result from massive stars, are observed in the thin disk and along the spiral arms of spiral galaxies. SN Ia result from older stars and occur throughout the Galaxy. Surveys of supernova remnants (SNRs) in the Milky Way (Case and Bhattacharya, 1998), pulsars in the Milky Way (Yusifov and Küçük, 2004), and SNe in nearby galaxies (van den Bergh, 1997) help to constrain the present radial distribution of SNe in the Milky Way. SNRs have the disadvantage that it is not easy to determine the types of supernovae that formed them.

The following is a simple parameterization of the radial surface density distribution of SNe in the Milky Way Galaxy:

$$f(r) = (A_1 + A_2) \left\{ \left( \frac{r}{r_o} \right)^\alpha \exp\left[ -\beta \frac{r}{r_o} \right] \right\} + A_3 \frac{6.5}{(r+2)^2} \tag{1}$$

where $A_1 = 3.55$, $A_2 = 0.745$, $A_3 = 0.00225$, $r_o = 8.5$ kpc, $\alpha = 4$, $\beta = 6.8$. $A_2$ and $A_3$ correspond to the SN Ia distribution, and $A_1$ corresponds to the SN II distribution. The first term and its $\alpha$ and $\beta$ constants are from Yusifov and Küçük's parameterization of the pulsar Galactic radial distribution. Since SNe Ia are observed to be more centrally concentrated in spiral galaxies than SNe II (van den Bergh, 1997), the $A_3$ term was included to account for SNe Ia in the nuclear bulge (its form was borrowed from Eq. 7 of Timmes *et al.*, 1995). SNe Ia also occur in the disk, requiring the $A_2$ term; the ratio of $A_2$ to $A_1$ was set to 21%, which is the SN Ia to SN II ratio in late-type spirals (Capellaro). According to Capellaro, SNe Ia are about 28% as frequent as SNe II in galaxies resembling the Milky Way; the ratios of the other $A$ constants were adjusted so that this SN Ia to SN II ratio is satisfied when integrated over the area of the Milky Way (from $r = 0$ to 20 kpc). Finally, all three $A$ constants were scaled so Eq. 1 gives the present Milky Way SN rate when integrated over the full disk (2.5 SNe century$^{-1}$). The fractional contributions from Eq. 1 from SNe II, disk SNe Ia, and bulge SNe Ia are 78%, 16%, and 6%, respectively. Equation 1 peaks at 5 kpc, where the surface density of supernovae is twice the value in the solar neighborhood. While the relative number of SN Ia is small in late-type spiral galaxies, each is more luminous than the typical SN II. Thus, the energy contribution from SNe Ia is not as insignificant as their relative rate implies. For an independent derivation of the SN distribution in the Milky Way, see Ferrière (2001).

Gehrels *et al.* (2003) explored the effects on Earth's atmosphere of the gamma ray and cosmic ray radiation from a nearby SN II. The gamma ray radiation resulting from the decay of $^{56}$Co lasts a few hundred days, while the elevated cosmic ray flux can last thousands of years. They ran their models for 20 years and found that a SN would have to occur within 8 pc for the UV radiation at Earth's surface to be at least doubled and estimate a rate of about 1.5 SNe Gyr$^{-1}$. They did not explore the biological effects of the secondary particle radiation produced in the atmosphere. Thus, their calculations should underestimate the important biological effects of a nearby SN II. There is also uncertainty regarding the effects of local magnetic fields on the propagation of the cosmic rays and also whether the effects of Earth entering the SNR.

GRBs are rare, very short-lived (~10s), very luminous (~$10^{51}$ ergs s$^{-1}$) events that produce most of their luminous energy in gamma rays with energies between 100 keV and 1 MeV (there is also some evidence that they produce TeV gamma rays). Scalo and Wheeler (2002) calculated that the GRB rate in the Milky Way is about one per $\geq$ 30,000 SN Ib/c. This rate depends on the assumed degree of collimation of the GRB radiation, GRB evolution with redshift, and the properties of the GRB progenitor (e.g., minimum mass star to explode as a SN Ib/c). There is still much room for improvement in our knowledge of these quantities.

Still, number of GRBs that affect Earth is not dependent on the degree of collimation, only on the observed rate; only those GRBs that we can see will affect us. The gamma ray photons from a GRB cannot reach the ground for planets with thick atmospheres like Earth's. The incoming photons are first downgraded to X-rays via Compton scattering, and then the X-rays are absorbed and generate photoelectrons. The energetic electrons then collide with oxygen and nitrogen atoms, exciting them and causing ultraviolet emission, which makes it to the surface. Scalo and Wheeler estimate that GRBs aimed at a planet with a thin atmosphere can do significant damage to the DNA of eukaryotes from as far as 14 kpc (nearly twice the distance to the Galactic center);

the corresponding distance for prokaryotes is 1.4 kpc. The critical distance for significant UV production at the surface of Earth is about 11 kpc. They estimate that this occurs once every 2-4 Myrs. The very short duration of the photon radiation burst on the surface of a planet resulting from a GRB implies that only life on one hemisphere of the planet will suffer its direct effects.

Melott *et al.* (2004) and Thomas *et al.* (2005) also consider the possible long-term damaging effects (a few years duration) of a GRB's photons on a planet's atmosphere, including ozone destruction, global cooling, and acid rain. They estimated that a GRB within 3 kpc of Earth can cause significant damage to its ozone layer, and that such an event should occur every 170 Myrs. They also suggested that a GRB might have caused the late Ordovician mass extinction.

While it has not been demonstrated by direct observation, it is likely that GRBs also generate collimated jets of energetic particles (Waxman 2004). Dar *et al.* (1998) and Dar and De Rújula (2002) consider the effects on the biosphere of particle jets from GRBs. These include atmospheric muons, radioactive spallation products, and ozone destruction. The muons can penetrate deep underwater and underground. The duration of the cosmic ray irradiation is expected to be ~2 days, long enough to cover all longitudes (but not necessarily all latitudes). Dar and De Rújula estimate that a GRB at the Galactic center aimed at Earth would produce a lethal dose of atmospheric muons to eukaryotes. The surface area exposed to muons would be maximal for a GRB that occurs above a planet's equator. Given the flattened distribution of stars in the Milky Way, a planet with an equator that coincides with the Galactic Equator will thus suffer the greatest extinctions.

The number of GRBs that affect Earth with their cosmic rays should be greater than those directly observed by their gamma rays. The cosmic rays emitted by a GRB with energies below the TeV range will be at least partially deflected from a straight-line trajectory by the Milky Way's magnetic fields. Thus, the effective opening angle of a GRB's cosmic ray jet should be greater for the lower energy cosmic rays.

Assuming the typical GRB results from the collapse of a very massive star, then the distribution of GRBs in the Milky Way should closely follow that of the pulsars. If, instead, they result from the mergers of binary compact objects, their distribution will be more spread out. Thus, as with SNe, GRBs should be more frequent inside the solar circle and their rate should be declining with time.

On December 27, 2004 a giant flare from the SGR1806-20 was observed with several space-based gamma ray detectors; it was the most powerful SGR flare observed to date (Hurley *et al*. 2005). Only four SGRs are known: three in the Milky Way and one in the Large Magellanic Cloud. The previous largest SGR flares were on March 5, 1979 and August 27, 1998 with energies between $10^{44}$ and $10^{45}$ ergs. The energy of the 2004 event was closer to the upper end of this range (Yamazaki et al., 2005).

A giant SGR flare likely results when a magnetar, neutron stars with extremely strong magnetic fields (up to $10^{15}$ G at the surface), experiences a sudden shift in their magnetic fields. Yamazaki *et al*. argue that the radiation from the 2004 event was collimated with an opening angle of about 0.2 radians. Thus, as is the case for GRBs, giant SGR flares should be more frequent than we observe them to be.

The gamma ray energy of the 2004 SGR flare was about six orders of magnitude less than that of the typical GRB. Assuming a SGR event of comparable luminosity is observed in the Milky Way once every 10 years, the total gamma flux received at Earth from SGRs should be comparable to that from GRBs. And, the effects on Earth would be similar. One possible difference would be that giant flares from a given SGR could repeat with intervals of decades or centuries.

An active Galactic Nucleus (AGN) outburst (or a nuclear starburst) is another type of transient radiation event. The Milky Way's nucleus is presently in a relatively inactive state, but there is strong evidence that a 2.6 million solar mass black hole resides there (Morris *et al.*, 1999); it is among the smaller black holes of those detected in nearby galaxies. The Milky Way Galaxy's nuclear black hole has grown over time by accreting gas and disrupted stars. While it is accreting, the black hole's accretion disk emits electromagnetic and particle radiation.

High-resolution observations indicate that all massive galaxies have a supermassive nuclear black hole (Miller *et al.,* 2003; Marconi *et al*., 2004). When active, such black holes are observed in the bulges of galaxies as AGNs (called Seyfert galaxies). The fraction of galaxies observed in the nearby universe with AGNs is related to the duty cycles of their black holes. Assuming the AGN–nuclear black hole paradigm is correct, then the larger the observed fraction of AGNs, the larger the average duty cycle. Large duty cycles are possible if AGNs are long-lived and/or frequent. Miller *et al*. studied the distribution of AGNs in the nearby universe and concluded that about 40% of massive galaxies have an AGN. They conclude from this that the typical AGN lifetime is about 2 x $10^8$ years, or that the typical AGN has burst 40 times over the 5.7 x $10^8$ years covered by their survey. Marconi *et al.* modeled the growth of nuclear black holes during AGN phases, and found that duty cycles have declined over the history of the universe and that they are larger for smaller black holes.

If these numbers can be applied to the recent history of the Milky Way Galaxy's nuclear black hole, then it should have been in an active state about 40% of the past half billion years. The luminosity of its nucleus in an active state would be about $10^{44}$ ergs s$^{-1}$ in 2-10 keV X-rays (Gursky and Schwartz, 1977). This energy range alone would generate the energy of a typical supernova in less than a year. Above about 5 keV, there is relatively little attenuation by the interstellar medium towards the Galactic center, but the absorption rises steeply towards lower energies. The total interstellar extinction towards the Galactic center is also a sensitive function of the distance from the Galactic mid-plane; planets located near the mid-plane will be the most protected from ionizing photons produced by an AGN. At Earth, the X-ray flux would be 130 erg m$^{-2}$ s$^{-1}$, assuming no intervening absorption. This is about 20 times the typical flux from the Sun in the same energy band and is comparable to the peak flux of an M-class X-ray solar flare. Including absorption would make the Galactic center X-ray flux comparable to that of the Sun's average value. Thus, the X-ray emission from an AGN outburst would probably not be very important for life on Earth, but it probably would be for planets within 1-2 kpc of the Galactic center. In addition, BL Lacertae objects and flat-spectrum radio quasars are observed have gamma ray luminosities up to $10^{49}$ erg s$^{-1}$ and $10^{50}$ erg s$^{-1}$, respectively (Hartman *et al.* 1999). Whether the Milky Way was ever in such states is another question.

Clarke (1981) argued that the particle radiation from an AGN outburst would be much more important to life than its ionizing photon emission. He calculated, assuming no propagation energy losses, that particle radiation fluxes at Earth would increase by a factor of ~100, causing significant damage to the ozone layer and increased radiation at the surface. This estimate needs to be checked with new calculations taking into account charged particle propagation along the Milky Way's magnetic field lines.

## 3.2.3 THE METALLICITY FACTOR

In order to model the GHZ, it is important to understand the spatial gradients and temporal evolution of the gas metallicity in the Milky Way. Galactic chemical evolution models are an important part of this, but they need to be checked against observations. Recent measurements of the disk radial metallicity gradient average near –0.07 dex kpc$^{-1}$ (e.g., Chen $et\ al.$, 2003). The metallicity of the gas in the disk is increasing with time; observations of thin disk G dwarfs in the solar neighborhood show that it is increasing at a rate of about 0.035 dex Gyr$^{-1}$, but there is still some controversy concerning the source of the scatter in metallicity at a particular age (see discussions in Gonzalez $et\ al.$, 2001a, Nordström $et\ al.$ 2004, and Pont and Eyer 2004). Observations also show that the radial gradient is increasing more slowly in the inner Galaxy, causing it to become shallower. Maciel $et\ al.$ (2005) find from recently published data that the flattening rate over the past 8 Gyrs has been in the range of 0.005 to 0.010 dex kpc$^{-1}$ Gyr$^{-1}$, with the present value being about -0.06 dex kpc$^{-1}$.

Knowledge of the metallicity gradient is critical to understanding the GHZ, as the initial gas metallicity of a cloud determines the properties of the terrestrial and giant planets that form from it. The incidence of Doppler-detected giant planets around nearby Sun-like stars is now known to be very sensitive to the host star's metallicity. It rises steeply from about 3% at solar metallicity to 25% at 0.3 dex (a factor of two) above solar metallicity (Santos $et\ al.$ 2004, 2005; Fischer and Valenti 2005). The lower metallicity bound determined from observations of nearby stars is about -0.55; in addition, surveys for transiting planets in the globular cluster 47 Tucanae ([Fe/H] = -0.76) have yielded null results (Gilliland $et\ al.$ 2000; Weldrake $et\ al.$ 2005).

Contamination of the host star's atmosphere by accreted solid protoplanetary disk material, once proposed to explain the correlation between stellar metallicity and the presence of giant planets, now appears to be at best a minor process (Gonzalez, 2003; Santos $et\ al.$ 2004). Two mechanisms are still consistent with the observations: planet migration and a metallicity dependence on giant planet formation. If the inward migration of a giant planet in a disk depends on metallicity, then it will be easier to discover giant planets with the Doppler method around metal-rich stars. Presently, the evidence for a correlation between the orbital periods of giant planets and metallicity is weak (Sozzetti, 2004). Therefore, the best explanation for the correlation between metallicity and the presence of giant planets is that giant planets are more likely to form around metal-rich stars. How the Solar System fits into this picture is still unsettled, but it is beginning to appear that is not typical (Beer $et\ al.$ 2004). The Sun is more metal-poor than most stars with Doppler-detected giant planets, and the orbit of Jupiter is less eccentric than most of those with orbital periods beyond a few days.

Ida and Lin (2004) and Kornet *et al.* (2005) have independently explored the metallicity dependence of giant planet formation with simulations based on the core-accretion scenario. They succeeded in accounting for the observed metallicity dependence of the incidence of giant planets. Since the processes in the early phases of giant planet formation apply also to terrestrial planet formation (prior to the gas accretion phase), studies like Ida and Lin's should give us a handle on the metallicity dependence of terrestrial planet formation as well. Gonzalez *et al.* (2001a) argued that the mass of a typical terrestrial planet should depend on the initial metallicity raised to the 1.5 power, while Lineweaver (2001) assumed (without justification) that the dependence should be linear. Clearly, additional theoretical research is required to better constrain this relation, and the planned Kepler mission (Borucki *et al.* 2003) promises to provide observations of terrestrial planet transits.

Arguably, the number of asteroids in a planetary system should also depend on the initial metallicity, given that asteroids are made of metals. While this is obviously true, the number of asteroids that remain in the asteroid belt after the planets form is also very sensitive to the properties of the closest giant planet to them. Thus, while asteroids do not form around very metal-poor stars, the number of asteroids that survive in a system is probably a highly non-linear function of metallicity. A good case can also be made that the number of comets in a system depends on metallicity (see below).

3.2.4 COMETS

The major comet reservoirs in the Solar System reside beyond the orbits of Neptune and Pluto (reviewed by Stern, 2003). Three reservoirs are typically recognized (listed with heliocentric distances): the Kuiper belt (30 to 1,000 AU), the inner Oort cloud (1,000 to 20,000 AU), and the outer Oort cloud (20,000 to 50,000 AU). There is direct observational support for the existence of the Kuiper belt and indirect evidence for the Oort cloud (Levison *et al.*, 2001). The Kuiper belt has about $7 \times 10^9$ comets, while the Oort cloud has about $10^{12}$ comets (Stern); the inner Oort cloud has about five times as many comets as the outer Oort cloud.

The properties of the Oort comet cloud around a given planetary system depend, in part, on the properties of its giant planets and the initial metallicity of its birth cloud. Presumably, a planetary system forming from an initially more metal-rich cloud will form a more populated Oort cloud, but this needs to be confirmed with self-consistent simulations that include the metallicity dependence of giant planet formation. Given the high sensitivity of giant planet formation to metallicity, it seems likely that this is a reasonable assumption. Granting this and the known Galactic metallicity disk gradient, planetary systems born in the inner Galaxy should start with more populous Oort clouds. The subsequent history of interaction between an Oort cloud and its Galactic environment is also critical to understanding the potential threats from comets.

The rate of nearby stellar encounters with the Oort cloud leading to perturbations of comet orbits is given by:

$$N = \pi D^2 \upsilon \rho \qquad (2)$$

where $D$ is a minimum interaction distance, $\upsilon$ is the velocity of the Sun relative to the perturbing star, and $\rho$ is the local stellar volume number density (Garcia-Sanchez et al., 1999). Therefore, for a given encounter distance and relative stellar velocity, the encounter rate will be proportional to the stellar volume number density. The magnitude of the perturbation on a comet goes as $\upsilon^{-1}$. However, the individual stellar velocity impulses must be added in quadrature, given their random distribution, so the net perturbation on the Oort cloud comets by passing stars will be proportional to $\upsilon^{-0.5}$ (Weissman, 1996). Thus, regions of the Milky Way with higher encounter velocities will lead to smaller perturbations of the Oort cloud comets, all else being equal. The typical relative velocity can be equated to the stellar velocity dispersion. Lewis and Freeman (1989) have determined the velocity dispersion of stars in the Milky Way's disk as a function of Galactocentric distance from observations of disk K giants (see their Figures 10 and 12). For instance, the velocity dispersion at 4 kpc is 68% greater than the local value.

Simple axisymmetric mass models of the Milky Way can be employed to explore how the comet threat varies with location. Dehnen and Binney (1998), for example, present analytical expressions for the axisymmetric mass density of the stellar and interstellar components as functions of Galactocentric distance, $R$, and distance from the mid-plane, $z$. From their expressions the stellar volume mass density can be calculated at any location relative to the local value (neglecting spiral arms, the inner bar, and other fine structure). Figure 2 shows the axisymmetric mass density at the mid-plane of the Galaxy between 0.1 and 20 kpc (calculated from the bulge and disk parameters of model 2 of Dehnen and Binney). The mass densities of gas and stars rise steeply toward the Galactic center. At the mid-plane the disk and bulge stars dominate the mass density inside of $R = 6$ kpc, while the disk gas dominates the outer Galaxy. If variations of the stellar mass function with location are neglected, then the relative stellar volume number density can be equated to the relative stellar volume mass density. Clearly, the stellar volume number density is a steep function of $R$. This implies, via Eq. 2, that stellar impulses on comets will also depend strongly on $R$. For example, taking into account only the variation in stellar density at mid-plane, the number of comets perturbed by nearby stellar impulses would be 5.3 times greater at $R = 4$ kpc compared to its value near the Sun's position at $R = 8$.

**[Figure 2]**

Hut and Tremaine (1985) have estimated that the number of comets perturbed by GMCs is similar to that perturbed by stars. The radial gradient of GMCs can be approximated by the gas mass density gradient determined from Dehnen and Binney's model; GMC density is 2.3 times greater at $R = 4$ kpc than at 8 kpc.

It is interesting to note that the Sun's Oort cloud is currently experiencing nearly the maximum perturbations compared to other times in its Galactic orbit. It is near the mid-plane and near the periapsis of its ~250 Myr orbit around the Galactic center. Weissman (1996) notes that this configuration enhances the perturbations of the Oort cloud comets by a factor near 2.5 compared to the average value. This should be taken into account when determining empirical estimates of the comet rate, as the presently observed rate does not correspond to the long-term average rate.

Since a planetary system in the inner Galaxy experiences more frequent perturbations from stars, GMCs and stronger tides, its comet reservoirs are depleted more rapidly; their "half-lives" are shorter. The history of perturbations of comet reservoirs around a planetary system at a given value of $R$ must be known if the relative comet threats are to be compared. It could be the case that a system's Oort cloud is depleted so rapidly a few billion years after forming that today it possesses too few comets to pose a threat. Hut and Tremaine, neglecting the Galactic tides, calculated the half-life of the Sun's outer Oort cloud due to GMC and stellar perturbations, finding ~3 Gyrs for each. Heisler (1990) considered only stellar and Galactic tidal perturbations, deriving a half-life for the outer Oort cloud of about 4.5 Gyr. Duncan *et al.* (1987), who included only stellar perturbations, found a half-life of 5 Gyr for the inner plus outer Oort clouds. An average half-life of 4 Gyrs for the complete Oort cloud is probably a reasonable overall compromise of these estimates.

The half-life of the Oort cloud due to stellar perturbations has a different functional dependence than that due to GMC perturbations (Hut and Tremaine). The half-life due to stellar perturbations is linearly proportional to the mean encounter velocity, but that due to GMCs is not dependent on velocity. Also, the half-life due to stellar perturbations goes as $a^{-1}$ while those due to GMCs goes as $a^{-1.5}$, where $a$ is the semi-major axis of a comet's orbit. The half-life due to Galactic tides should go as $a^{-1}$. Since the Galactic tides dominate, the half-life should go approximately as $a^{-1}$. Given this, the half-life of inner Oort cloud comets with an average $a$ value 20% of the outer cloud comets should have a half-life five times greater.

The half-life of an Oort cloud varies with location in the Milky Way. Interstellar gas metallicity, stellar velocity dispersion, stellar density, GMC density, and Galactic tides all vary with location and all affect the Oort cloud comets. To first order, the initial number of comets in the Oort cloud can be approximated as being proportional to $10^{[Fe/H]}$, where [Fe/H] is the logarithmic number abundance of Fe relative to the solar value. Given that the Galactic tide is the dominant perturber of the Sun's comets, the half-life of each reservoir can be assumed to be inversely proportional to the total mass volume density (stars plus gas) at mid-plane and only weakly dependent on the stellar velocity dispersion. Therefore, the half-life of the Oort cloud of another Sun-like star is approximately:

$$\tau_O = 4 \left( \frac{\rho}{\rho_s} \right)^{-1} \sqrt{\frac{v}{v_s}} \quad \text{Gyrs} \tag{3}$$

where the $s$ subscripts refer to the values of the quantities at the Sun's present position.

The relative rate of Oort comet perturbations from stars, GMCs, and the disk tide (neglecting the radial tides) is:

$$\frac{F}{F_s} = \frac{N_O}{N_{Os}} \left( 0.1 \frac{\rho_{star}}{\rho_{stars}} \sqrt{\frac{v_s}{v}} + 0.1 \frac{\rho_{ISM}}{\rho_{ISMs}} + 0.8 \frac{\rho_{all}}{\rho_{alls}} \right) \tag{4}$$

where the three terms enclosed by parentheses correspond to stellar, GMC, and disk tide (which depends on the total mass density) perturbations, respectively. The $N_O$ coefficient is equal to the number of comets in a star's Oort cloud; it will depend on the initial metallicity and the history of perturbations. The contribution from the Kuiper belt has been neglected. The numerical constants multiplying each of these three terms correspond to relative contributions at the Sun's present position of 10%, 10%, and 80% for the stars, GMCs, and tides, respectively; the specific values of these constants are not meant to be precise. According to Eq. 4, moving the Solar System inward to $R = 4$ kpc, for example, would increase the comet flux by a factor of 3.4, which is dominated by the Galactic tide term. Taking into account the higher metallicity for a system formed at 4 kpc would increase the comet flux by another factor of 2. A more complete treatment would have to take into account also the evolution of the number of Oort cloud comets around a star. Thus, for example, an inner Galaxy Oort cloud would initially produce more comet showers than the Solar System's Oort cloud, but it would be depleted more quickly.

The properties of an Oort cloud also depend on the mass of its host star. If the properties of giant planets depend on the mass of their host stars, then so will the properties of the Oort cloud. In addition, an Oort cloud like ours around a less massive star should have a smaller half-life, due to the weaker gravitational binding energy.

In addition to the threat from comets residing in the Sun's gravitational domain, there is also the threat from interstellar comets. They can be grouped into two distinct types: 1) free floating comets lost from Oort clouds around other stars, and 2) comets gravitationally bound to other stars. In the following, interstellar comets of the first type are designated as IS1 and those of the second type as IS2.

Weissman (1996) reviewed the threat posed to Earth by IS1 comets and estimated their number density in the solar neighborhood between $10^{12}$ to $3 \times 10^{13}$ IS1 comets $pc^{-3}$, assuming the following: 1) the Solar System has ejected $4 \times 10^{13}$ to $9 \times 10^{14}$ comets over its history, 2) all stars are formed with Oort clouds like ours, and 3) a mean space volume of 12 $pc^3$ per star. Since an interstellar comet will be on a hyperbolic trajectory upon passage near the Sun, it should be easy to distinguish it from the abundant comets with elliptical and parabolic orbits that are gravitationally bound to the Sun. To date, no comets with hyperbolic orbits have been observed. McGlynn and Chapman (1989) calculated that six comets with hyperbolic orbits should have been observed if the local density of IS1 comets is $10^{13}$ $pc^{-3}$; Sekanina (1976) had calculated an upper limit of 4 x $10^{12}$ IS1 comets pc$^{-3}$. These observational limits correspond to the low end of Weissman's predictions.

The probability that an interstellar comet collides with Earth is much greater than just the geometrical cross section of Earth amplified by the Sun's gravity; it depends on the cross section of the orbits of the giant planets, which can capture an interstellar comet into an orbit that brings it into the inner planets region. Taking this factor into account and assuming a local density of IS1 comets of $10^{12}$ pc$^{-3}$, Zheng and Valtonen (1999) calculated that Earth should have collided with about 100 such comets over its history, or one every 40 million years.

Type IS2 comets are not included in the observational limits, because they are episodic. Presently, there are no stars sufficiently close to the Sun for their bound comets to pass near enough the Sun for us to see them. Type IS2 comets only pose threats to us when their host star passes within about one-third of a parsec of the Sun. Thus, a nearby star passage will threaten Earth both from its comets and from the perturbed comets around the Sun. Taking an average heliocentric distance in the Sun's inner Oort cloud (10,000 AUs) and the present number of comets there ($\sim 10^{12}$), the average number density is 2 x $10^{15}$ pc$^{-3}$ (or, about 1 comet per 5 cubic AUs). This is at least three orders of magnitude greater than the number density of IS1 comets in the solar neighborhood.

The probabililiy of impact from interstellar comets should vary with Galactic location. The three key factors are: metallicity, stellar density, and stellar velocity dispersion. The higher metallicity in the inner Galaxy should result in more interstellar comets there. The inner Galaxy should be populated by a higher density of IS1 comets resulting from more frequent star-star and star-GMC encounters and stronger Galactic tides. On the other hand, the larger stellar velocity dispersion in the inner Galaxy will reduce the cross-section for comet captures. The more efficient stripping of comets from their Oort clouds in the inner Galaxy will reduce the importance of type IS2 comet collisions relative to those of type IS1; the accelerated stripping of comets from a star's Oort cloud in the inner Galaxy is at least partly compensated by the increased number of interstellar comets.

In a preliminary work, Masi *et al.* (2003) calculated the effects of the Galactic radial tide on outer cloud Oort comets at different locations in the disk. They confirmed that the perihelia of the outer Oort cloud comets are decreased to the region of the terrestrial planets much more effectively in the inner Galaxy and that at $R = 2$ kpc tidal stripping of the outer Oort comets becomes severe.

### 3.2.5 GIANT MOLECULAR CLOUDS

While a planetary system is traversing interstellar space, it will occasionally encounter a GMC (and more often, lower density clouds), and the probability of encounter is increased when it is crossing a spiral arm. Talbot and Newman (1977) calculated that the Solar System should encounter an average density GMC (~330 H atoms $cm^{-3}$) every 100 Myrs and a dense GMC (~2 x $10^3$ H atoms $cm^{-3}$) every Gyr. The biologically significant effects of such an encounter are varied. These could include comet showers, exposure to a greater flux of cosmic ray particles, glaciations and more nearby SNe.

Begelman and Rees (1976) first noted that passage of the Sun through interstellar clouds with densities of at least $10^2$-$10^3$ H atoms $cm^{-3}$ are sufficient to push the heliopause inside Earth's orbit. This would leave Earth exposed to interstellar matter. Scherer *et al.* (2002) noted that the shrinking of the heliopause will also eliminate the solar modulation of the cosmic ray flux at Earth and expose it to a higher flux, possibly leading to more low clouds due to the GCR-cloud link noted above. Florinski *et al.* (2003) determine that the GCR flux at Earth would be enhanced by a factor of 1.5-3 just by a cloud with a hydrogen density of 8.5 $cm^{-3}$, which is about 30 times the present local interstellar gas density. The cosmic ray flux within a GMC would be much greater due both to recent SNe and to the longer cosmic ray diffusion time.

Yeghikyan and Fahr (2004a,b) modeled the passage of the Solar System through a dense interstellar cloud ($10^3$ $cm^{-3}$), confirming that the heliopause is pushed in to the region of the terrestrial planets. In such a situation the interstellar matter interacts directly with Earth's atmosphere. They studied the consequences of accretion of hydrogen, disputing the earlier claim of Yabushita and Allen (1997) that passage through a GMC core would deplete all the oxygen in the atmosphere. They found, instead, that the ozone in the upper atmosphere is depleted, and Earth is cooled by about 1 degree C, speculating that it might cause an ice age.

Pavlov *et al.* (2005a) noted two effects neglected in previous studies that should significantly increase the effects to the biosphere of a cloud passage. First, the flux of anomalous cosmic rays (generated from interstellar neutrals) would increase much more than the GCR flux. Second, during passage through a typical cloud (lasting ~ 1 Myr) there would be one or two magnetic field reversals, during which time Earth's atmosphere would remain unprotected from cosmic rays at all latitudes. Pavlov *et al*. calculated that passage through a cloud with a density of 150 H atoms cm$^{-3}$ would decrease the ozone column by 40% globally and up to 80% near the poles.

Pavlov *et al*. (2005b) studied the possible effects of dust accumulation in Earth's atmosphere during passage through a GMC. They concluded that such an event could produce global snowball glaciations and that less dense clouds could still produce moderate ice ages.

Of course, a planet closer to the outer edge of the CHZ would feel the effects of clouds passages more intensely and more often, as would a planet with a weak or absent magnetic field. Thus, among the terrestrial planets in the Solar System, Mars should have been subjected to the most severe interactions with interstellar clouds. This is another reason that the CHZ definition should be expanded beyond merely the maintenance of liquid surface water.

How would the effects from passages through GMCs vary with location in the Galaxy? First, GMC encounters should be more frequent in the inner Galaxy due to the higher density of GMCs there. Second, the encounter velocities should be greater in the inner Galaxy, making it more likely that the heliopause will be pushed back to the vicinity of the terrestrial planets, even for less dense interstellar clouds. In addition, at high encounter velocity, the energy deposited by interstellar dust impacting a planet's atmosphere might become an important factor in atmospheric loss, but this needs to be explored quantitatively.

### 3.2.6 GALACTIC DYNAMICS

Galactic dynamics is arguably the most complex aspect of the GHZ. While the Sun's original birth orbit in the Galactic disk cannot be determined, there is little doubt that it has changed significantly since it formed. Older stars in the disk have larger velocity dispersions (see Gonzalez, 1999 and references therein). Stars form in relatively circular orbits, and over time they experience gravitational perturbations that make their orbits more eccentric and send them farther from the disk mid-plane. Thus, older stars tend to pass through the mid-plane at higher velocity and traverse a greater range of radial distances from the Galactic nucleus.

Since stars migrate in the disk, the present metallicity of interstellar gas in the star's vicinity, corrected for the time of its formation, will not be representative of the star's metallicity. The disk radial metallicity gradient gives us the metallicity of a star at the time of its formation, but the star will later wander to a region with a higher or lower gas metallicity. As a result, the GHZ has fuzzy boundaries. Since stellar migration in the disk is a stochastic process, the problem of defining the GHZ should be treated with Monte Carlo simulations.

The spiral arms are important structures for determining the boundaries of the GHZ. They contain most of the GMCs and SNe II in the Galaxy. The interarm regions have less star formation activity, but the star density there is only slightly less than that in the arms at the same Galactocentric radius. Within the framework of the density wave hypothesis, the spiral arms rotate about the nucleus of the Milky Way like a solid body, with a constant angular frequency (Lin *et al.,* 1969). Assuming the Milky Way's spiral arm pattern can be maintained for at least several Gyrs (see Lepine *et al.*, 2001), then present surveys of the spiral arms can be used to conduct long-term simulations of stars' motions with respect to them.

Stars at the corotation circle will orbit the Galactic center with the same period as the spiral arm pattern. Thus, the interval between spiral arm crossings is longest for a star in a circular orbit at the corotation circle. However, Lepine *et al.* (2003) show that a star near the corotation circle experiences resonant perturbations with the spiral density waves that cause it to wander in the radial direction by 2-3 kpc in less than a Gyr. This would imply that spiral arm crossings are minimized, instead, at a moderate distance from the corotation circle. If the Sun is indeed very close to the corotation circle, as some studies indicate, then it is a surprise that its orbit has such a small eccentricity. Its small eccentricity implies that the Sun is not very near the corotation circle. There is a great need to settle the question of the Sun's distance from the corotation circle.

The interarm spacing at the Sun's location is 2.5 ± 0.3 kpc (Vallée, 2002). The Sun is presently located 0.20 kpc inside its mean Galactic orbital radius and about 0.14 kpc from its perigalactic radius (Sellwood and Binney, 2002). It is 0.9 ± 0.4 kpc from the Sagittarius arm (Vallée). Thus, the Sun is presently safe from radial excursions into either the inner or outer arms; this is not the case if the Sun is far from the corotation circle.

Shaviv (2003) and Shaviv and Vezier (2003) claim to have found a link between spiral arm crossings and long-term variations in Earth's climate, especially extensive glaciations. At the core of their thesis is the GCR-cloud link. They reconstruct the historical GCR flux from meteorite exposure ages and the concurrent temperature variations from ancient calcite shells. They compare these data to the calculated cosmic ray flux fluctuations resulting from spiral arm crossings (varying from about 35% to 140% of the present value). These data are open to interpretation, and their thesis has been criticized for multiple reasons (Rahmstorf *et al*. 2004; Royer *et al*. 2004). Also, the strength of the correlation depends sensitively on the precise value of the Sun's position relative to the corotation circle. Nevertheless, the correlations are intriguing and deserve additional attention.

De la Fuente Marcos and de la Fuente Marcos (2004) calculated the variation in star formation in the solar neighborhood from open cluster data. They found a strong correlation between the peaks in the star formation rate and the glaciation events of the past 1 Gyr. While this confirms part of Shaviv's thesis, the variation in star formation offers an alternative explanation for the cosmic ray flux variations at Earth than spiral arm crossings.

In summary, the evolution of the GHZ is determined primarily by Galactic chemical evolution. In a given region of the Galaxy, Earth-size terrestrial planets are unlikely to form until the interstellar gas metallicity reaches a value close to solar metallicity. Survival of life depends on the distribution of radiation hazards and comet perturbers. Threats to life increase toward the center of the Galaxy and decrease with time. The greatest uncertainty about the GHZ concerns stellar dynamics and how a given star's orbit interacts with the spiral arms.

3.3 THE COSMIC HABITABLE AGE

Progress in refining the CHA will come primarily from improvements in our understanding of the evolution of the cosmic star formation rate. Star formation, in turn, determines the evolution of the average supernova rate, AGN activity, and gas phase metallicity in galaxies. Many of the same studies relevant to the GHZ can also be applied to the CHA.

If all galaxies were just like the Milky Way, then the GHZ could just be applied to other galaxies. But, they aren't; there is great variation in their properties. Galaxies differ in their Hubble types (elliptical, spiral, or irregular), metallicities, luminosities, masses, and environments. Some of these properties correlate with each other. For example, mass correlates with luminosity. Metallicity also correlates with luminosity, in the sense that more luminous galaxies are more metal-rich (the so-called L-Z relation). Lamarielle *et al*. (2004), for example, determined a log (O/H) to absolute magnitude slope of about –0.26 from both blue and red-band observations in the large 2dFGRS spectroscopic survey; see Kuzio de Naray *et al*. (2004) for a summary of other recent studies of the L-Z relation. This trend can be seen even among the small number of galaxies in the local group. For example, the Small and Large Magellanic Clouds have metallicities about –0.7 and –0.4 dex relative to the gas in the solar neighborhood, respectively.

Calura and Matteucci (2004) calculated the evolution of the production of metals over the history of the universe. They determined that the present mean metallicity in galaxies is close to the solar value. This is consistent with the fact that the Milky Way Galaxy is among the 1-2% most luminous galaxies in the local universe. The inner Galaxy experienced more metal enrichment over its history and has a higher mass density than the solar neighborhood. Thus, the mass-weighted average of the Galaxy's metallicity is larger than the solar value. In addition, luminosity-weighted metallicity averages for galaxies tend to be smaller than mass-weighted ones, because the bright red giants in a galaxy tend to be more metal-poor.

Even though the average metallicity of local galactic matter is now close to solar, there were many metal-rich stars formed within the first 2-3 Gyrs after the Big Bang. The metals first built up quickly in the inner regions of (now) massive galaxies. Presumably, these metal-rich stars have been accompanied by planets since that time. However, like the juvenile Milky Way Galaxy, these were also the regions with the most dangerous radiation hazards, including supernovae, GRBs, and intense AGN activity.

Galaxy evolution is not the same everywhere. Spheroids are much more common in clusters than in small groups, like the Local Group. Spiral galaxy disks tend to be stripped of gas more efficiently in dense cluster environments, and star formation is suppressed as a result (Vogt *et al*., 2004). Galaxy collisions are also less frequent in sparse groups.

Galaxy mergers also shape galaxies and alter their star formation rates. Major mergers can temporarily increase the star formation rates throughout each of the involved galaxies (becoming starburst galaxies) and possibly consume most of the gas. Thus, some fraction of merging galaxies become ellipticals following exhaustion of their gas, with a greatly reduced star formation rate (and thus a reduced planet formation rate). Mergers can also feed fresh gas into any supermassive black holes in their nuclei, causing AGN outbursts. Mergers were frequent during the first few Gyrs after the Big Bang, but they continue to the present at a low rate. For example, Hammer *et al*. (2005) argue that about three-quarters of intermediate size spirals have experienced mergers within the last 8 Gyrs.

Surveys of supernovae and their remnants tell us not only about the present supernovae distribution and rate in our Galaxy, they also reveal how these quantities vary among the various types of galaxies. Surveys of SNRs in large nearby galaxies (Matonick and Fesen, 1997; Sasaki, 2004) show that their radial scale lengths are generally similar (about 30% their disk radii), and many have peaks at 20% to 40% of their disk radii (the Milky Way being such a case). Some spirals have different radial SNR distributions. The starburst spiral galaxy NGC 6946, for example, has a sharp peak of the SNR density at its nucleus. Particularly helpful are studies like Capellaro (2004), which catalogs the observed rates of all SN types for all types of galaxies. Such studies reveal that SNe II do not occur in ellipticals, while they predominate in late-type spirals, where the overall SN rate is greater.

With the star formation rate continuing to slow, the mean metallicity will increase ever more slowly. Many galaxies presently below solar metallicity will eventually build up enough metals to form Earth-size terrestrial planets. The time in the history of the universe when a particular region in a given galaxy reaches this critical stage is important. Too early, and the radiation environment may be too intense for life. There is also a limit at late times. As the star formation rate declines, the production of the long-lived geologically important radioisotopes cannot keep up with their decay in the ISM. Earth-size terrestrial planets forming in the future will have less radiogenic heating (Gonzalez *et al.*, 2001a). Of course, increasing the size of a planet can compensate for this deficit, but then all the processes dependent on planet size discussed above will need to be taken into account (such as reduction in surface relief).

Massive stars will become rarer, which means that the rate of GRBs, SGR flares and SNe will decrease. AGN activity will also decline. Galaxies will continue to recede from each other. On the other hand, radioisotope concentrations will continue to decline, G dwarfs will become rarer and stellar galactic orbits will become hotter. Adams and Laughlin (1997) speculated on the fate of the universe in the far distant future, after nucleons decay and black holes evaporate. While these are extreme events, there is no question that the universe will change drastically as it evolves beyond its present state. The changes will take it ever further from the conditions we know are consistent with life (especially complex life).

Based on the evolution of the global star formation rate, the CHA encompasses at least the last 3.5 Gyrs (the time since the origin of life on Earth) and probably the last 5 Gyrs and the next 10 Gyrs. The future limit is based primarily on the half-lives of the geologically important radioisotopes and the decline in formation of G dwarfs. It should be noted that this future habitability limit of the universe is not a hard limit; some locations should be habitable perhaps even 20 Gyrs from now. Still, this is brief compared to the possible future history of the universe.

Interestingly, there appears to be a convergence of the timescales of several important processes noted above that permits life to flourish on Earth within this brief window. The following life-relevant processes have comparable timescales (1-10 Gyrs): the nuclear evolutionary timescale of the Sun; the rate of decline of the Sun's activity; the mean half-life of the geologically important radioisotopes; the loss rate of Earth's volatiles; the recession rate of the Moon and the slowdown of Earth's rotation; the evolution of the star formation rate in the Galaxy; and the expansion rate of the universe. Within an anthropic framework these coincidences could be interpreted as implying that all these processes and their particular timescales were necessary for our existence as observers.

## 3.4 FILLING IN THE GAPS

Several important processes fall in the gaps between the CHZ and GHZ and also between the GHZ and CHA as they have been described above. The birth cluster of a habitable system is the most glaring gap between the CHZ and GHZ. Star clusters in the Milky Way Galaxy range from dense metal-poor globular clusters to sparse metal-rich Galactic clusters. The properties of a star cluster depend on the time and location of its formation in the Galaxy. However, considerable variety exists among clusters in the solar neighborhood and among globular clusters in the halo. Star cluster formation needs to be incorporated into Galactic chemical evolution models with the appropriate stochastic variations to account for the observed range of their properties. Once formed, the dynamical evolution of a planetary system must be followed in the context of the Galactic potential and the stellar, gas, and dust environment. Since the dynamical evolution of a star in the Galaxy has a strong stochastic component, it will be necessary to employ Monte Carlo methods to treat it properly.

In an analogous way to the star cluster gap between the CHZ and GHZ, there is a galaxy cluster gap between the GHZ and CHA. The local galactic environment of the Milky Way Galaxy is sparser than many other groups and clusters, but it is denser than that of an isolated galaxy. How does the habitability of a galaxy depend on its local environment? This is another question that needs attention.

The GHZ is presently defined only for the Milky Way Galaxy. Extension to the broader universe will also require generalization of the GHZ to other kinds of galaxies. In particular, the stellar dynamics in other Hubble types are very different from those in the disk of the Milky Way Galaxy. What does the GHZ look like for very different Hubble types, such as ellipticals and irregulars?

As a way of summary, some of the complexity of the many habitability factors is presented in highly schematic form in Figure 3. The many interrelationships are represented by the links between the boxes. The overall problem of habitable zones is highly nonlinear. It will require considerable computational power to perform the required Monte Carlo simulations and long temporal integrations. Its solution will require much more advancement in astrophysics, geophysics, climatology, and biology.

**[Figure 3]**

## 4. Conclusion

The present status of research both directly and indirectly related to the three types of habitable zones: CHZ, GHZ, and CHA has been reviewed. While the most progress has been made on the CHZ, there are still surprisingly many processes that have yet to be incorporated into its definition. The traditional definition of the CHZ needs to be expanded beyond the narrow focus on insolation evolution and atmospheric processes. A self-consistent picture of the CHZ is not possible without including the other diverse and interdependent factors.

The GHZ and CHA are newer concepts and, not surprisingly, have received less attention. Both can benefit from ongoing surveys of the Milky Way Galaxy and nearby and distant galaxies. Some of the most relevant results from these surveys include SN statistics, luminosity functions of galaxies and clusters of galaxies, and the evolution of the star formation rate. Continued theoretical studies of the effects of SNe and AGN outbursts on life, stellar dynamics in the Milky Way and in other galaxies, and Galactic chemical evolution will also aid in refining the GHZ and CHA.

As broad as they are, even these three habitable zones still leave gaps. To paint a complete picture of the evolution of the habitability of the whole universe, it will be necessary to merge these presently disparate concepts. Thus, many opportunities remain for astrobiologists.

## Acknowledgments

The author thanks John J. Matese for helpful advice on Oort cloud comet perturbations. Also, the two anonymous referees provided very helpful comments.

Table 1. Planetary initial parameters used for the N-body cases (columns 1-5) and the resultant mean and maximum eccentricities for the Earth (columns 6,7). The mass of Earth is in units of Earth masses, and that of Jupiter is in Jupiter masses. The semi-major axes are in units of AUs. Each case is a 15 Myr integration of Earth, Venus and the four giant planets, except 2* and 15*, which are 45 Myr. Case 1 is the nominal case for the present Solar System.

| Case | $M_{Earth}$ | $a_{Earth}$ | $M_{Jup}$ | $a_{Jup}$ | $e_{Jup}$ | mean $e_{Earth}$ | max $e_{Earth}$ |
|------|-------------|-------------|-----------|-----------|-----------|------------------|-----------------|
| 1    | 1           | 1           | 1         | 5.2       | 0.048     | 0.025            | 0.045           |
| 2    | 1           | 1           | 1         | 5.2       | 0.097     | 0.046            | 0.080           |
| 2*   | 1           | 1           | 1         | 5.2       | 0.097     | 0.047            | 0.080           |
| 3    | 1           | 1           | 1         | 5.2       | 0.146     | 0.045            | 0.120           |
| 4    | 1           | 1           | 1         | 5.2       | 0.194     | Unstable         | Unstable        |
| 5    | 1           | 1           | 0.2       | 5.2       | 0.048     | 0.030            | 0.089           |
| 6    | 1           | 1           | 0.5       | 5.2       | 0.048     | 0.053            | 0.091           |
| 7    | 1           | 1           | 2         | 5.2       | 0.048     | 0.019            | 0.037           |
| 8    | 1           | 1           | 3         | 5.2       | 0.048     | 0.016            | 0.034           |
| 9    | 1           | 1           | 4         | 5.2       | 0.048     | 0.016            | 0.033           |
| 10   | 1           | 1           | 5         | 5.2       | 0.048     | 0.018            | 0.039           |
| 11   | 1           | 1           | 1         | 4.2       | 0.048     | 0.107            | 0.502           |
| 12   | 1           | 1           | 1         | 3.2       | 0.048     | 0.023            | 0.049           |
| 13   | 1           | 1           | 1         | 2.7       | 0.048     | 0.026            | 0.055           |
| 14   | 0.1         | 1           | 1         | 5.2       | 0.048     | 0.061            | 0.129           |
| 15   | 0.5         | 1           | 1         | 5.2       | 0.048     | 0.028            | 0.051           |
| 15*  | 0.5         | 1           | 1         | 5.2       | 0.048     | 0.028            | 0.052           |
| 16   | 2           | 1           | 1         | 5.2       | 0.048     | 0.023            | 0.040           |
| 17   | 1           | 1.5         | 1         | 5.2       | 0.048     | 0.026            | 0.047           |

FIGURE CAPTIONS

*Figure 1*. Distribution of the perihelia of nearly 2 x 10$^5$ asteroids and 400 comets. The aphelia of the four terrestrial planets are also shown. Asteroids are about 2.5 times more common near Mars compared to Earth. Note, the Kirkwood gaps are only visible on histograms that plot semi-major axis; plotting perihelion distance, instead, smears out the gaps. The comet perihelia distribution peaks near Mars. (The vertical axis of the top diagram is split for clarity.) Asteroid data are from the Minor Planet Center and include all objects in their database larger than about half a kilometer (to minimize detection bias), as of June 21, 2003. Comet data are from JPL's DASTCOM database as of June 21, 2003. Both sets of data are restricted to asteroids and comets with known orbital elements.

*Figure 2*. Mass density of stars and gas at the mid-plane of the Milky Way Galaxy; both the bulge and disk components are included. This is based on model 2 of Dehnen and Binney (1998). The Sun's position is shown as an open circle on the curve showing the total density.

*Figure 3*. Highly schematic diagram showing the many interrelationships among the important habitability factors. The factors with the ellipses have significant stochastic components. The factors for terrestrial planets continue along the bottom of the figure, as they did not all fit along the left column.

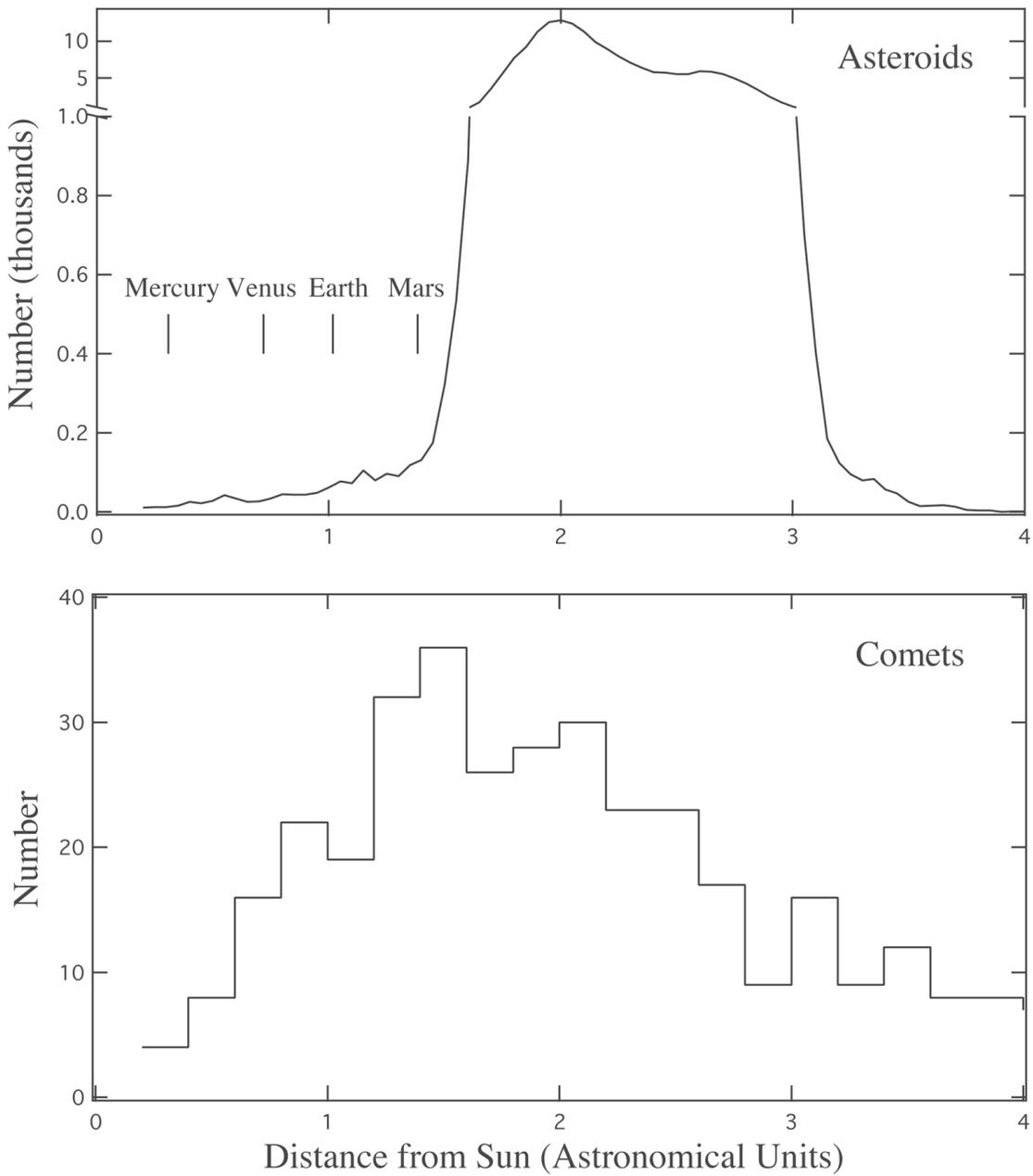

Figure 1: asteroids and comets

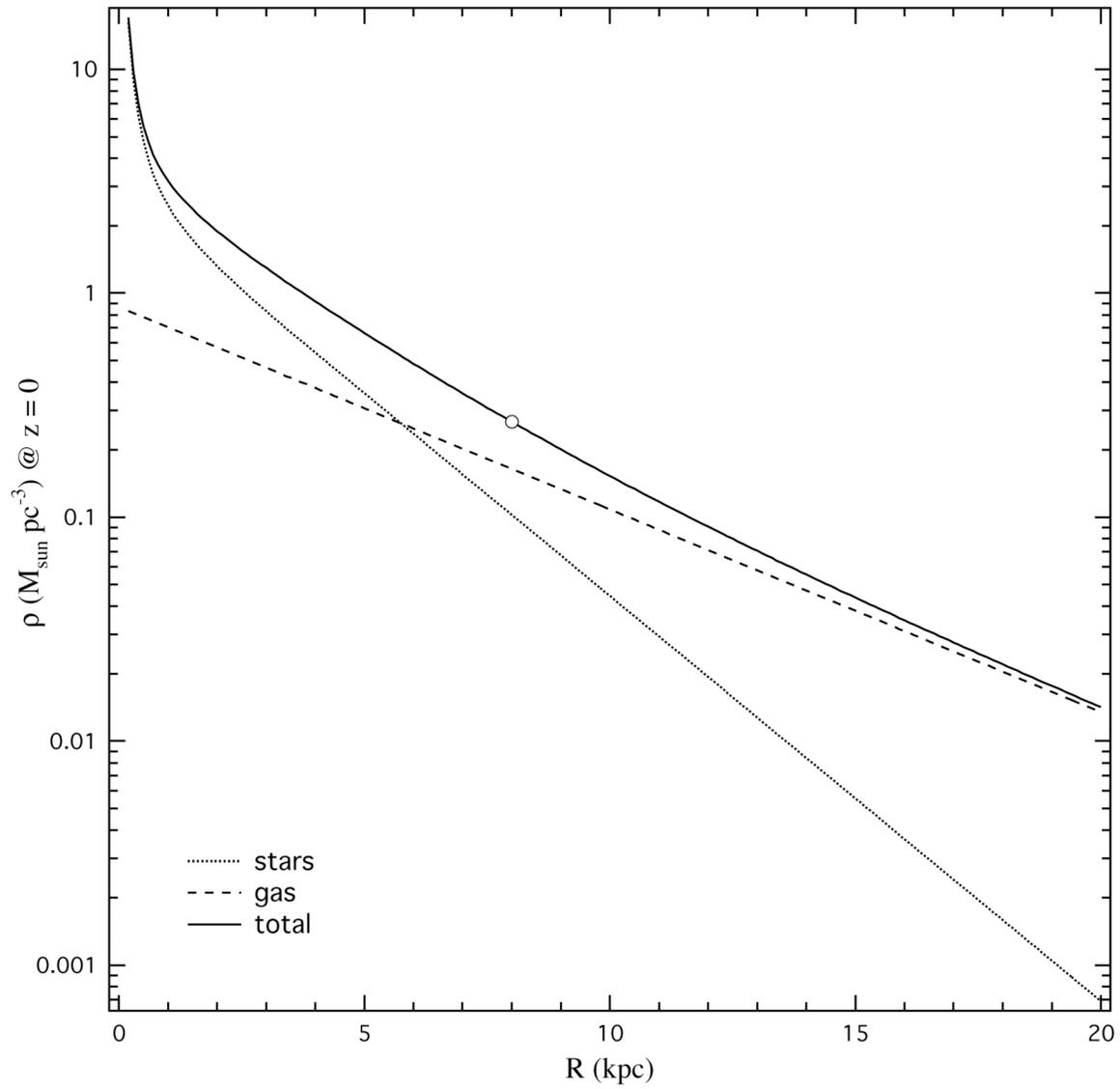

Figure 2: Galactic mass distribution

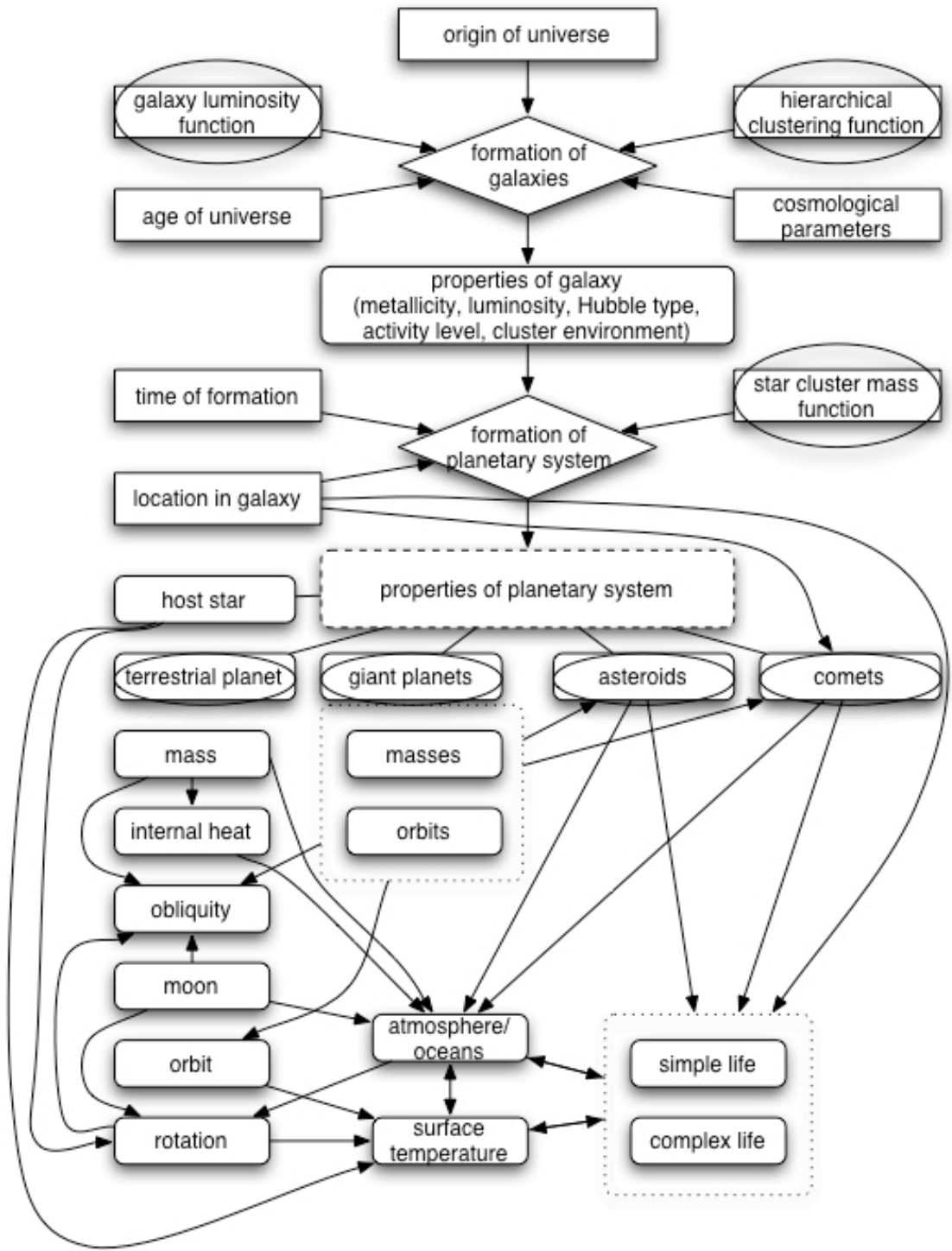

Figure 3: schematic of habitability